\documentclass[12pt,preprint]{aastex}

\usepackage{natbib}

\newcommand\cmt{{\rm cm^{-3}}}

\newcommand\K{{\rm K}}

\newcommand\Msun{{\rm\,M_\odot}}

\newcommand\kms{{\rm km\, s^{-1}}}
\newcommand\pc{{\rm\,pc}}
\newcommand\kpc{{\rm kpc}}
\newcommand\simgt{\lower.5ex\hbox{$\; \buildrel > \over \sim \;$}}
\newcommand\simlt{\lower.5ex\hbox{$\; \buildrel < \over \sim \;$}}

\shorttitle{Stratified ISM Disks with MRI-Driven Turbulence}
\shortauthors{Piontek \& Ostriker}

\begin{document}


\title{Models of Vertically Stratified Two-Phase ISM Disks 
with MRI-Driven Turbulence}


\author{Robert A. Piontek\altaffilmark{1,2}}
  \email{rpiontek@aip.de}

\author{Eve C. Ostriker\altaffilmark{1}
\email{ostriker@astro.umd.edu}}

\altaffiltext{1}{Department of Astronomy, University of Maryland,
    College Park, MD  20742-2421}

\altaffiltext{2}{Astrophysikalisches Institut Potsdam, An der
Sternwarte 16, D-14482 Potsdam, Germany}



\begin{abstract}    
  We have performed time-dependent numerical simulations of the
  interstellar medium (ISM) which account for galactic shear and
  magnetic fields, vertical gravity, and a radiative cooling function
  for atomic gas.  This allows us to study the magnetorotational
  instability (MRI) in cloudy, vertically-stratified disks.  As in
  previous unstratified models, we find that thermal instability
  interacts with MRI-driven turbulence and galactic shear to produce a
  network of cold, dense, filamentary clouds embedded in a warm
  diffuse ambient medium.  This structure strongly resembles the
  morphology of HI gas observed in the 21 cm line.  There is
  significant thermally-unstable gas, but the density and temperature
  distributions retain the twin peaks of the classical two-phase ISM.
   Independent of the total gas surface density and vertical gravity
   levels adopted, the midplane ratios of thermal
   to magnetic pressure are $\beta =0.3-0.6$, when the mean vertical
   magnetic field is $0.26 \ \mu G$.  
  We analyze the vertical
  distributions of density and various pressure terms, and address what
  supports the ISM vertically.  All models become
  differentially stratified by temperature; 
  only when the cold mass fraction is small does turbulent mixing 
  maintain a large cold-medium scale height.  Turbulent
  velocities of the cold gas also increase as the cold mass fraction
  decreases, but are generally low ($\sim 1-3\ \kms$) near the
  midplane; they increase to $>5\ \kms$ at high $|z|$.  Turbulent amplitudes
  are higher in the warm gas. 
  The central thermal pressure is similar 
  for all models even though the total weight varies
   by a factor 7 for a range of imposed vertical gravity; in higher
   gravity models the increased weight is supported by increased
   magnetic pressure gradients.   Approximate vertical equilibrium holds
   for all models.  
  Finally, we argue
  that in the outer parts of galactic disks, MRI is likely able to
  prevent the development of self-gravitating instabilities and hence
  suppress star formation, even if cold gas is present.
\end{abstract}




\section{Introduction}

The classical picture of the ISM began to take shape with Field's work
on thermal instability (1965), and subsequently the realization that
for realistic heating and cooling functions the atomic ISM could exist
in two distinct stable phases in pressure equilibrium
\citep{1969ApJ...155L.149F}. \citet{1977ApJ...218..148M} argued that
supernovae (SNe) transform the ISM, allowing for variations in
pressure, driven turbulence, and a hot component formed by SNe blast
waves that overrun a significant fraction of the volume.  Many elements of 
this ``three-phase'' model are still being scrutinized by both
observers and theoreticians today.  For understanding the ISM's
thermodynamics, perhaps the most important development in more recent
years is that both observations and simulations have found large
fractions of gas to exist at temperatures which are thought to be
thermally unstable.  This calls into question whether or not a two or
three phase model of the ISM is valid, and more generally how thermal
and dynamical processes interact in the ISM.

Surveys of the HI 21cm line have shown that the ISM is very turbulent.
Typical turbulent velocities are found to be approximately $7\ \kms$
\citep{2003ApJ...586.1067H, 2004JApA...25..185M}.  In the traditional
picture of the ISM, the source of turbulence is attributed to SNe
\citep{1978ppim.book.....S}.  Many simulations have been performed
which explore the effects of SNe on the ISM, with increasingly complex
methods used for incorporating star formation (e.g.
\citet{{2004A&A...425..899D},{2005A&A...436..585D},{2005MNRAS.356..737S}}).
Most simulations have looked only at SNe rates equal to or above what
is thought to be typical of the Milky Way.  However,
\cite{2005ApJ...630..238D} find that for SNe rates lower than half of
the mean Galactic rate, the velocity dispersions fall short by a
factor of 2--3 compared to typical observed values. Since low
turbulence disks containing cold gas may be susceptible to violent
gravitational instabilities, this suggests a source of (kinetic and/or
magnetic) turbulence other than SNe may be needed to self-consistently
explain ISM conditions in the outer Milky Way -- and other galaxies -- where
there is little star formation.

Various extragalactic observations have also implied that sources of
turbulence other than SNe may be present.  In external galaxies, the
ISM appears to be turbulent regardless of whether an active star
forming region is nearby. In NGC 1058, spiral arm and interarm regions
have indistinguishable HI velocity dispersions
\citep{1990ApJ...352..522D}), and turbulence levels in outer galaxies
are comparable to those in inner galaxies, even though star formation
rates drop off steeply \citep{1999AJ....118.2172V}.

These results have led us to explore another of the physical mechanisms which
has potential to drive turbulence in the ISM. Among candidate mechanisms,
perhaps the most viable in the outer parts of galaxies (where self-gravity is
weak) is the magneto-rotational instability (MRI) \citep{1969A&A.....1..388F,
  1991ApJ...376..214B, 1991ApJ...376..223H}.  The
MRI may also contribute significantly in inner galaxy regions. In the past
decade the MRI has been studied extensively in the context of accretion disks
surrounding compact objects and protostars (e.g. 
\citet{2000prpl.conf..589S,2003ARA&A..41..555B}).  The MRI generates
turbulent velocities and amplifies magnetic fields in magnetized, shearing
disk systems, which leads to the transport of angular momentum outward through
the disk, allowing matter to accrete toward the center. Galactic disks
meet the basic requirements for the MRI to be present: a moderate 
magnetic field
and decreasing angular velocity with increasing radius.
\cite{1999ApJ...511..660S} suggested that turbulence in outer
disks may be driven by the MRI.

Previous single-phase simulations of MRI in the context of galactic
disks have been performed by \citet{2003ApJ...599.1157K} (local
models) and \citet{2004A&A...423L..29D} (global models). We have
addressed the issue of MRI in multiphase gas with direct numerical
simulations, beginning with \citet{2004ApJ...601..905P} (hereafter
Paper I), in which we performed two dimensional computations in the
radial-vertical plane.  These models were local, and utilized a linear
galactic shear profile and shearing-periodic boundary conditions, a
radiative cooling function which allowed for a two phase medium, and
magnetic fields.  Our two dimensional models were extended to three
dimensions in \citet{2005ApJ...629..849P} (hereafter Paper II),
allowing us to study late time evolution, as two dimensional
simulations do not (and cannot) yield saturated state turbulence. We
found in Paper II that the saturated state velocities can reach as
high as $8\ \kms$ in two-phase simulations with a mean density $\bar
n=0.2\ \cmt$, which is comparable to the low values found in the outer
Galaxy.  Turbulent velocity amplitudes were found to scale with mean
density as $\delta v\propto \bar n^{-0.77}$, while the saturated-state
magnetic field strength was independent of $\bar n$, with plasma
$\beta\equiv P_{\rm th}/P_{\rm mag}\approx 0.5$ within each phase.

In this paper we extend the simulations of Paper II to include
vertical gravity, thus allowing for stratification of the disk to
develop. This enables us to address a longstanding question in studies
of the ISM, namely, how is the gas vertically supported against
gravity?  Potentially, vertical support can be provided by thermal
pressure, kinetic (ram) pressure, and magnetic stresses (and
indirectly by cosmic-ray pressure).  \citet{1990ApJ...365..544B}, 
for example, argue
that kinetic, magnetic, and cosmic ray terms contribute nearly equally
to the vertical support of the gas.  With a multiphase, cloudy medium,
however, the dynamics of vertical support must be quite complex, and
in particular the various components may be unequally supported and
hence become differentially stratified.  For atomic gas,
investigations in the 1970's indeed suggested that the cold HI
component of the disk is signficantly thinner than the warm HI component
\citep{1973A&A....25..253F,1975ApJ...198..281B}.  While an updated
observational evaluation of the differential HI stratification would
be very valuable, it is equally important to address the theoretical
issue with modern, numerical methods.  As we shall show with our
simulations, cold and warm gas become differentially stratified, with
high density cold clouds preferentially sinking towards the mid-plane.
We find that the degree of differential stratification depends on the
relative proportions of cold and warm gas.

In this work, we address a number of issues, many of which were not
possible with the models presented in Paper II.  What fraction of the
ISM is found to exist in each of the warm, cold, and thermally
unstable phases when turbulence is driven by the MRI?  What is the
vertical profile in a self-consistent turbulent system, and what
fraction of vertical support is provided by thermal, kinetic, and
magnetic stresses?  Can turbulence driven by the MRI provide the
necessary effective pressure to reproduce the observed scale height of
the galactic disk?  How do turbulent (kinetic and magnetic) amplitudes
of stratified models compare with previous unstratified simulations, in
particular for conditions that prevail in the outer, non-star-forming
regions of galactic disks?  To explore dependence on parameters, we
address these (and other) questions by performing three simulations
with fixed surface density which differ in the strength of gravity by
a factor of sixteen, and an additional outer-disk model with low
surface density and low gravity.

In \S \ref{numerics}, we describe our numerical method and the
parameters of our models.  In \S \ref{results} we present our results
and analysis.  In the final section we summarize and discuss the
implications of our results, and present concluding remarks.

\section{Numerical Methods and Model Parameters}
\label{numerics}

We solve the equations of ideal 
MHD in a local model representing sheared rotating
flows, with additional terms for radiative heating and cooling, heat
conduction, and gravity in the vertical direction:

\begin{equation}
\frac{\partial\rho}{\partial t}+ {\bf \nabla} \cdot (\rho
{\bf v}) = 0
\end{equation}

\begin{equation}
\frac{\partial{\bf v}}{\partial t}+
{\bf v}\cdot{\nabla}{\bf v}=-\frac{\nabla P}{\rho} + 
\frac{1}{4\pi\rho}({\nabla} \times {\bf B})
\times {\bf B}+ 2 q {\Omega}^{2}x\hat{x}-2{\bf \Omega}\times {\bf v} + {\bf g}_{ext}
\label{mom2a}
\end{equation}

\begin{equation}
\frac{\partial \mathcal{E} } { \partial t }
+{ \bf v}\cdot \nabla\mathcal{E} = 
-(\mathcal{E} + P)\nabla\cdot{\bf v}-\rho\mathcal{L}+
\nabla\cdot(\mathcal{K}\nabla T)
\end{equation}

\begin{equation}
\frac{\partial {\bf B}}{\partial t}=\nabla \times({\bf v}
\times {\bf B}).
\end{equation}



All symbols have their usual meanings.  Rotational shear is described
in terms of the dimensionless shear parameter $q\equiv-d \ln \Omega /
d \ln R$, where $q$ is set to one to model a flat galactic rotation
curve.  We model the gravitational force ${\bf g}_{ext}$ as a linear
function of height (assuming the gas scale height is smaller than most
of the total mass), 
\begin{equation}
{\bf g}_{ext}=-\tilde{g} z\hat{z}.
\end{equation}
The cooling
function, $\mathcal{L} = \rho \Lambda(\rho,T) - \Gamma$, is adopted
from \citet{2002ApJ...577..768S}, and is a piecewise power-law fit to
the data of \citet{1995ApJ...443..152W}.  The heating rate,
$\Gamma=0.015 \ {\rm ergs \ s^{-1}}$, is spatially constant, modeling
heating due primarily to the photo-electric effect of UV starlight on
small grains and PAHs.  This cooling function allows for two stable
phases of gas, warm and cold, to coexist in pressure equilibrium.  For
the adopted cooling curve parameters, the minimum and maximum
pressures for cold and warm stable equilibria, respectively, are
$P_{min, cold}/k=800\ \rm{K \ cm^{-3}}$ and $P_{max, warm}/k=3100\ \rm{K \
  cm^{-3}}$.  Based on the transition temperatures in
our adopted cooling curve, cold gas is defined to be below 141 $K$,
warm gas above 6102 $K$, and unstable gas between these two
temperatures.  The conduction coefficient is set so that we can
numerically resolve the length scales of thermal instability, with
$\mathcal{K}=1.03 \times 10^{7} \ \rm{ergs \ cm^{-1} \ K^{-1} \
  s^{-1}}$ (this is larger than the realistic level of conduction in
the ISM, but as discussed in Paper I the value does not affect our
late-time results).  Without conduction, TI would be most unstable at
the grid scale.

We use a modified version of the ZEUS MHD code
\citep{1992ApJS...80..753S,1992ApJS...80..791S} to integrate equations
(4.1) - (4.4) in time.  ZEUS is a finite difference, operator split,
time-explicit method for solving the equations of MHD. Shocks are
captured via an artificial viscosity.  Paper I gives a complete
description of our numerical implementation of heating, cooling, and
conduction, as well as code tests. The same basic methods were used in
Paper II, for a version of the code parallelized with MPI.  The
majority of the simulations presented here were run on the Thunderhead
cluster at Goddard Space Flight Center, while others were run on
the cluster at the Center for Theory and Computation at the University
of Maryland.

The primary difference between the numerical approach
here and that of Paper II is the addition of the gravity term in equation
(\ref{mom2a}), which leads to vertical density stratification.
 We have made use of a density floor in order to prevent the time
step from becoming prohibitively short, and also applied 
the Alfv\'{e}n limiter of
\citet{2000ApJ...534..398M}.  The density floor was set to $0.004 \ \cmt$ ,
and $c_{lim}=8$ in our high surface density runs.  For our outer galaxy model,
which has a larger vertical extent than other models, the density floor was
reduced to $0.0008 \ \cmt$.    The effect of the Alfv\'{e}n
limiter is essentially to limit the speed of MHD waves in a low density
environment.  

The use of a density floor has the undesired effect of increasing the
total mass contained in the simulation domain over time.  The total
mass in our simulations  increases by 60 -- 70\% over the course of the
whole run (10 orbits).  We attempted to assess the significance of
this by reducing the density floor.  This leads to a reduction in the
time step in order to satisfy the Courant condition, however, so that
the total number of numerical steps taken over a given physical time
increases; thus lowering the density floor had little effect on the net
increase in mass.  Limitation of computational resources prohibits us
from performing a comparison simulation without a density floor.  In
the following section we plot quantities such as mass fractions,
turbulent velocities, and pressure as a function of time.  These do
not show any significant trends, indicating that the increase in the
total mass due to the density floor is not seriously affecting our
results.  Furthermore, since the total fractional mass increase is
almost the same for all models, their relative global parameters are
the same as in the initial conditions.

Shearing-periodic boundary conditions are used in the radial ($\hat
x$) direction
\citep{1992ApJ...400..595H,1995ApJ...440..742H,1996ApJ...463..656S},
while periodic boundary conditions are used in the azimuthal ($\hat
y$) and
vertical ($\hat z$) directions.  Though the use of periodic boundary conditions
in the vertical direction is not ideal, it offers some advantages
compared to outflow boundary conditions, which we also explored.
Depending on the implementation of outflow boundary conditions, they
may or may not maintain the divergence free constraint of the magnetic
field.  Outflow boundaries can also create strong unphysical magnetic forces as
magnetic flux is advected through the boundary and field lines are
``cut'' \citep{1996ApJ...463..656S}.  Our simulations with outflow
boundaries also lost significant amounts of mass over the course of
the simulations.  However, we have verified that the important
features of our simulations are the same regardless of the choice of vertical
boundary conditions.

The simulations are initialized with an isothermal disk in vertical
gravitational equilibrium; thus the initial profiles are Gaussian.  The
parameter space explored by our models is summarized in Table \ref{table1}.  For our
standard model, the initial temperature of the disk is $2500 \ K$, the
mid-plane density is $n_0=0.85 \ \cmt$, and the scale height, $H$, is 150 pc.  The
total gas surface density is therefore $\Sigma=\sqrt{2\pi} \rho_0 H =10 \ 
M_\odot \ {\rm pc^{-2}}$.  To explore the effect of gravity we also performed
a high gravity model, as well as a low gravity model, respectively increasing
and decreasing $\tilde g$ by a factor 4. The values of the gravity constant
are $\tilde{g}=1.94 \times 10^{-31} {\rm s}^{-2}$, $7.76 \times 10^{-31} {\rm
  s}^{-2}$, and $3.10 \times 10^{-30} {\rm s}^{-2}$.  
To obtain the same initial mid-plane
density and scale height as the standard model, and hence the same value of
$\Sigma$, the initial disk temperatures in the low- and high-gravity
 models were set to $600 \ K$ and 
$10,000 \ K$, respectively.\footnote{After cooling is
  initialized at $t=4$ orbits the temperatures and pressures in these three
  models evolve to be roughly the same.}  We adopt $P_0/k=2000 \ K \ \cmt$ as
the unit of pressure in the code; we use this value of $P_0$ to normalize the
pressure in a number of the figures.  The central density and surface density
were chosen to be generally consistent with estimated ranges near the Solar
radius in the Galaxy.  The largest value of $\tilde{g}$ is comparable to the
estimated value near the Solar radius using $\tilde{g}=4\pi G \rho_{tot}$, for
$\rho_{tot}\approx 0.09 \ M_\odot \ {\rm pc^{-3}}$ the combined (stars + gas)
mid-plane density \citep{2000MNRAS.313..209H}.  The lowest gravity model is
representative of outer-galaxy conditions (see below).

The magnetic field is initially vertical,
with a plasma beta parameter $\beta=100$ 
for the standard
model.  In the high and low gravity runs $\beta$ was adjusted so that the
initial magnetic 
field strength was the same as in the standard model.  
\footnote{This initial vertical magnetic
field strength, $0.26\mu{\rm G}$, is comparable to the local estimated
vertical magnetic field of $0.2-0.3\mu{\rm G}$
\citep{1994A&A...288..759H}; the mean vertical magnetic
field is constant in our simulations, due to imposed boundary conditions.}
Random white
noise perturbations are added at the 1\% level to seed the MRI.  Cooling is
not turned on until $t = 4.0$ orbits, just as the MRI modes begin to become
non-linear.  Most simulations last approximately 10 orbits, with one orbit
set equal to 240 Myrs.  The size of the computational volume is $2H \ \times \ 2H
\ \times \ 6H$ (for $H= 150$pc the initial scale height), 
with 128 $\times$ 128 $\times$ 384 grid zones, giving a
resolution of 2.3 pc.

We performed an additional simulation with a lower initial surface density of
$\Sigma=6 \ M_\odot \ {\rm pc^{-2}}$, to explore how MRI would behave under
conditions more representative of outer galaxies.  The gravity constant is the
same as that of the low-gravity model described above (which is comparable to
the level that would be provided without a stellar disk).  
In this model the initial
scale height of the isothermal disk was increased to 300 pc, with a
corresponding initial temperature of $2500 \ K$.  The initial mid-plane
density was set to $0.26 \ \cmt$.  The initial magnetic field strength and
configuration were identical to the previously-described models.

A final set of five simulations were performed to assess the importance of two
important parameters: the initial magnetic field 
strength and the box size. These
models were all run at half the linear resolution of our standard model.  The
control simulation, BOX1, is otherwise identical to the standard model, and
serves as a point of comparison to assess the impact of reducing the
resolution.  In models BOX2 and BOX4 we increased the box size in both the X
and Y directions by a factor of 2 and 4, respectively, and also increased the
number of grid points by the same factor so the linear resolution is
unchanged.  
The final two models, MAG1 and MAG2, are identical to BOX1 and BOX2,
but have an initial magnetic field strength of $0.08 \mu{\rm G}$, or
$\beta=1000$.

In order to address the question of what supports the gas vertically in our
turbulent disk models, and how much of each phase is present, we need a
baseline for comparison.  For this comparison we have performed (two
dimensional) simulations which do not include magnetic fields, and therefore
do not develop MRI-driven turbulence.  Cooling is initialized at $t = 0$, and the gas quickly
separates into two stable phases.  The cold component settles to the
mid-plane, with the warm gas found above and below the cold gas in pressure
equilibrium with gravity. Both the warm and cold gas lie essentially
on top of the thermal equilibrium curve in the $P-\rho$ plane.  The warm and cold components
are separated by a thin layer of unstable gas, and there is little mixing
between the two stable phases.  We performed four non-turbulent simulations,
varying the initial temperature of the disk and ${\bf g}_{ext}$, as in the
turbulent models; all initial conditions other than $\beta$ are the same as
adopted in the corresponding turbulent models.

For all the non-turbulent models, the pressure at the
point where the gas makes a warm/cold transition is 
$P_{\rm trans}/k=1100-1200 \ K \ \cmt$ .  This is
close to the value $P_{\rm min, cold}/k=800 \ K \ \cmt$ which defines the
minimum on the cold branch of the thermal equilibrium curve.  Since the cold
medium has a very small scale height, the warm/cold transition occurs very
close to the midplane.  As a consequence, the warm medium has a total surface
density close to that of an isothermal layer with central pressure 
$P_{0, \rm warm}=P_{\rm trans}\simgt
P_{\rm min, cold}$; i.e.
\begin{equation}\label{Sigw}
\Sigma_{\rm warm} =\frac{ P_{0,\rm warm}}{c_{\rm warm}}
\left(\frac{2\pi}{\tilde{g} }\right)^{1/2}
=2.1\, M_\odot\ {\rm pc}^{-2} 
\left(\frac{P_{0,\rm warm}/k}{1000 \ {\rm K}\ {\rm cm}^{-3}}\right) 
\left(\frac{c_{\rm warm}}{8 {\ \kms } }\right)^{-1} 
\left(\frac{\tilde{g}}{ 10^{-30} {\rm s}^{-2} }\right)^{-1/2}, 
\end{equation}
where $c_{\rm warm}=8 {\ \kms } $ is the sound speed typical of the warm
medium near $P_{\rm min, cold}$.  The cold medium must contain the balance of 
the total surface density present, 
$\Sigma_{\rm cold}=\Sigma-\Sigma_{\rm warm}$.  Thus, for the 
$\Sigma =10\, M_\odot\, \pc^{-2}$ models with low, medium, and high gravity,
the predicted warm fractions for the non-turbulent two-layer case 
with $P_{0, \rm warm}= 1000 \ {\rm K}\ {\rm cm}^{-3}$ would be
0.48, 0.24, and 0.12, respectively.  For the 
$\Sigma =6\, M_\odot\, \pc^{-2}$ outer-disk model, the predicted warm fraction  
for a non-turbulent disk would be 0.79.

\section{Results}
\label{results} 
\subsection{Evolution}
From $t = 0$ to approximately $t = 4$ orbits the disk is in pressure
and gravitational equilibrium, with both heating and cooling disabled
until this time.  This method of initialization prevents the creation
of a very thin cold disk, as would occur if the cooling function were
initialized at $t = 0$.  During these first few orbits the modes of
the MRI begin to grow and strengthen from the small amplitude
perturbations imposed in the initial conditions.  At around $t = 4$
orbits the modes of the MRI begin to saturate due to nonlinear
interactions, at which point the cooling function is enabled.  The
disk then undergoes thermal instability and rapidly evolves into a
two-phase medium which is no longer in vertical gravitational equilibrium.  The
heavier cold clouds quickly sink towards the mid-plane, but kinetic
and magnetic turbulence driven by the MRI limits this
settling (somewhat).  Some of the largest turbulent amplitudes are seen during
this stage as the channel flow forms, and then quickly breaks up.  A
quasi-steady state is soon established, after which time the averaged
mass fractions, turbulent velocities, magnetic field strengths, and
other quantities are fluctuating, but remain roughly constant.  

\begin{figure}
\plotone{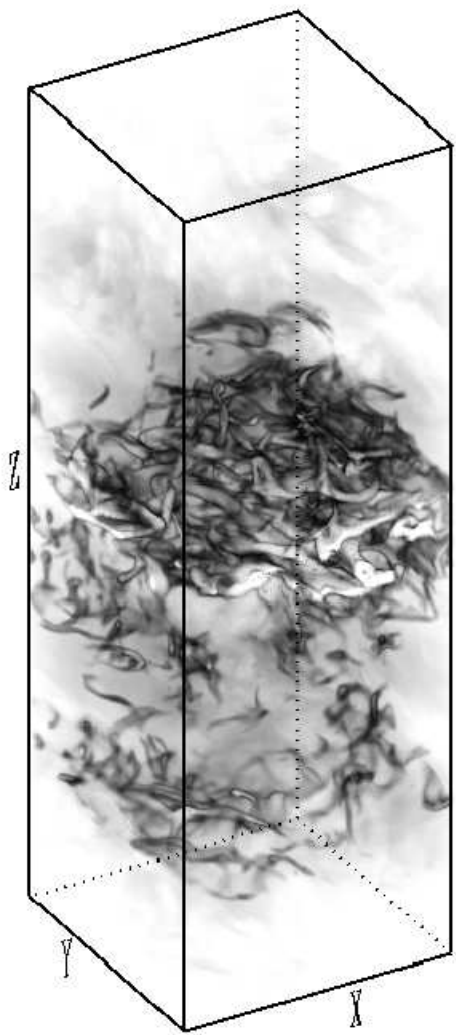}
\caption[Volume rendering of density for the standard gravity
run]{Volume rendering of density for the standard gravity run, at t=8
orbits.
\label{f1}}
\end{figure}

\begin{figure}
\plotone{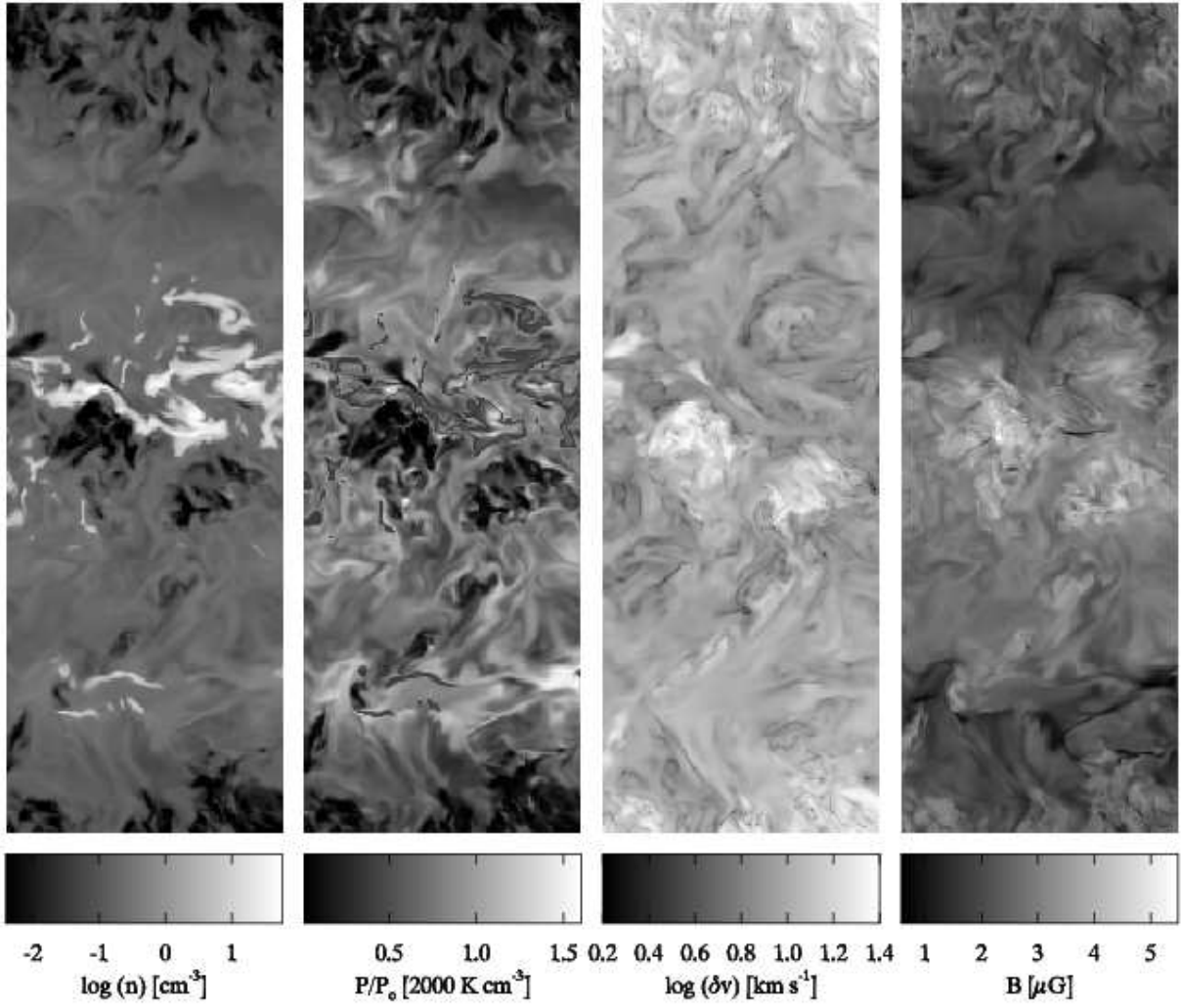}
\caption[Slices through the computational volume of the field
variables]{Slices through the computational volume of the field
variables for the standard model, at $t=8$ orbits.  From left to
right: density, thermal pressure, perturbed velocity, and magnetic
field strength.
\label{f2}}
\end{figure}

The MRI continues to drive turbulence throughout the duration of the
simulation.  In Figure \ref{f1} we show a volume rendering of
density for our standard model late in the simulation at $t = 8$
orbits. Though these simulations are stratified, the overall character
of these models evidently appears similar to those
presented in Paper II.  In particular, the high density clouds are
generally quite filamentary in character.  In Figure \ref{f2} we
show slices through the computational volume of the field variables:
density, pressure, velocity, and magnetic field.  Notice that while
most of the high density gas is located near the midplane, there are
still some high-density structures at large $|z|$.  Also evident in
the midplane are regions of quite low density and thermal pressure
where the magnetic fields and velocities are large.

The high-gravity and low-gravity models evolve similarly to the standard
model.  The evolution of our outer galaxy model differs from these 
in one important
respect.  When cooling is enabled at $t = 4$ orbits,
cold gas does not immediately form, because ambient pressures are not high
enough.
Only after the modes of the MRI have grown significantly are turbulence-induced
compressions able
to force gas into the cold, high density state.  Initially this occurs
throughout the simulation domain, at both low and high latitudes.  Over time,
however, the cold medium settles to the mid-plane.

In the following sections we analyze the turbulent velocities, magnetic
fields, pressure, and thermal structure of the gas.  We discuss the time
history of averaged quantities, as well as presenting probability distribution
functions of thermal and magnetic pressure, temperature, and density. In
addition, we analyze the vertical structure of our models, considering the
question of how material is supported against gravity.  Throughout, we make
comparisons among results of models with varying ${\bf g}_{ext}$.

\subsection{Turbulent velocities}

\begin{figure}
\plotone{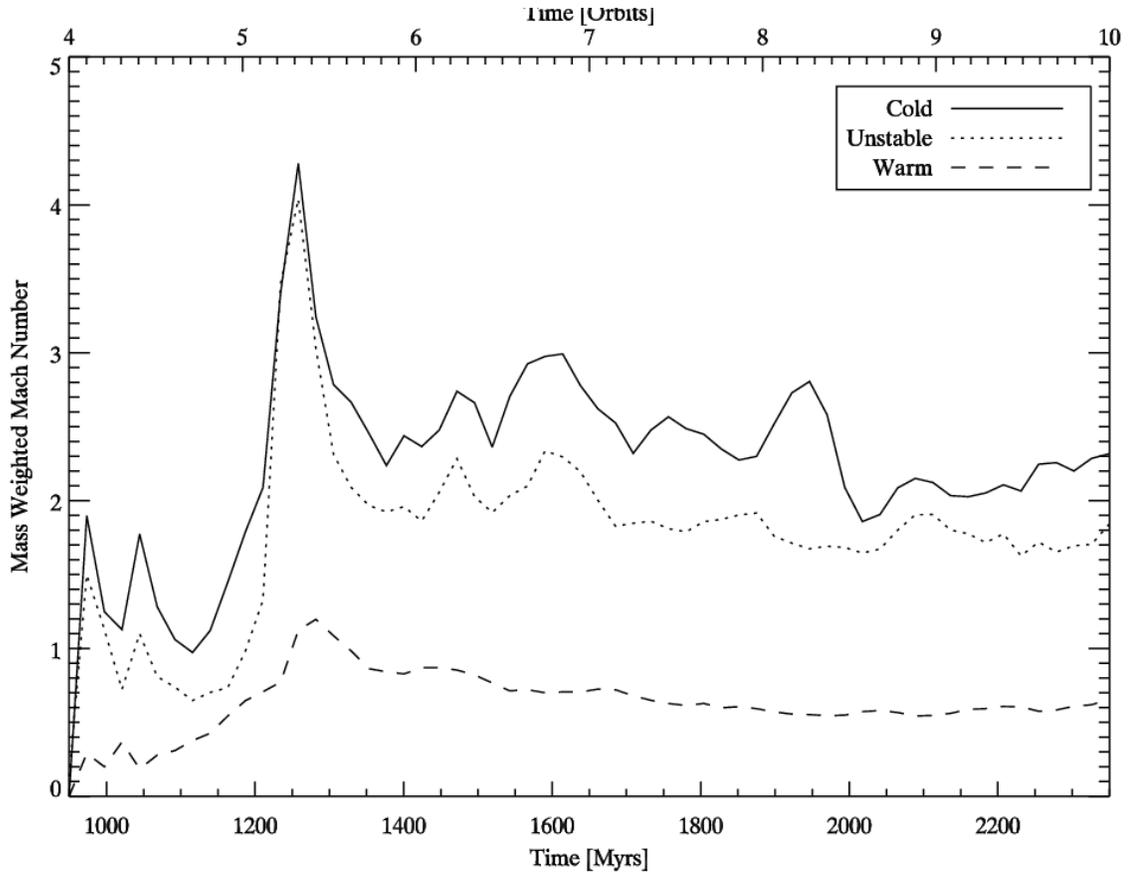}
\caption[Mass-weighted Mach numbers for the cold, unstable, and warm
phases of gas]{Mass-weighted Mach numbers for the cold, unstable, and
warm phases of gas in the standard run, plotted against time from
t=4--10 orbits.  The time-averaged Mach numbers over orbits 8--10 are
2.2, 1.7, and 0.6.
\label{f3}}
\end{figure}

\begin{figure}
\plotone{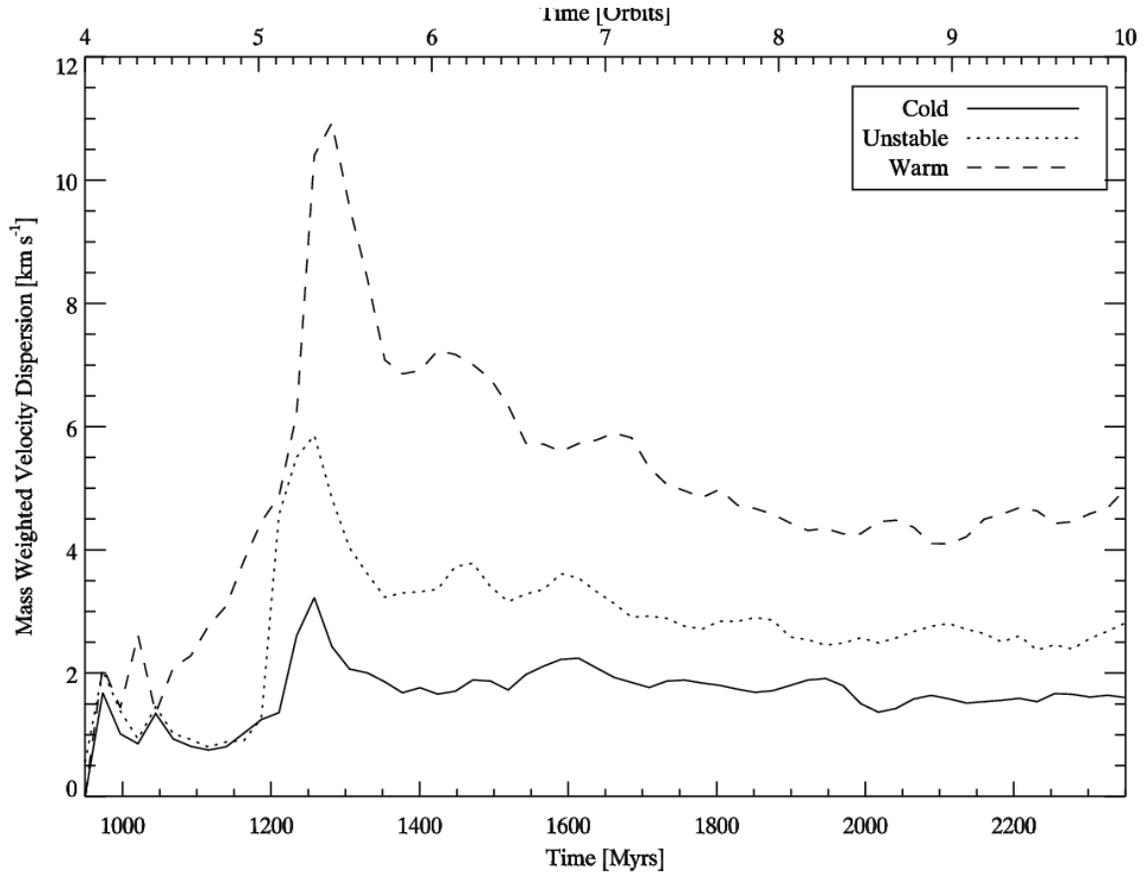}
\caption[Mass-weighted velocity dispersion for the cold, unstable, and
warm phases]{Mass-weighted velocity dispersion for the cold, unstable,
and warm phases for the standard model.  The late-time averaged values
for the warm, unstable, and cold phases are 4.4, 2.6, and 1.6 $\kms$.
\label{f4}}
\end{figure}

The mass-weighted RMS Mach numbers 
${\cal M} \equiv [\sum \rho (\delta v / c_s)^2 /\sum \rho]^{1/2}$ are plotted
in Figure \ref{f3}, from $t = 4-10$ orbits for the standard run
(cooling is disabled prior to $t=4$ orbits).  The
isothermal sound speed $c_s=(P/\rho)^{1/2}$ is computed individually
for each grid zone, and the galactic shear component is subtracted
from the azimuthal velocity as $\delta v_y=v_y + q\Omega x$ so that
$\delta v \equiv \sqrt{v_x^2 + (\delta v_y)^2 + v_z^2}$. The saturated
state Mach numbers for the warm, intermediate and cold phases of gas,
averaged over orbits 8--10, are 0.6, 1.7 and 2.2 for the standard
model.  
In Figure \ref{f4} we plot the corresponding mass-weighted
velocity dispersions for each of the three components; the time
averaged values are 4.4, 2.6, and 1.6 $\kms$, averaged over the same
interval.  The largest velocities (up to $11\ \kms$) are observed when
the MRI channel solution is strongest, at $t\sim 5.5$ orbits.

In the high gravity model, the respective mass-weighted velocity
dispersions are 6.5, 2.8, and 1.4 $\kms$, for the warm, intermediate,
and cold phases; in the low gravity model they are 4.9, 2.9, and 1.7
$\kms$; and the outer galaxy model they are 6.6, 3.3, and 2.5 $\kms$.
There is therefore little difference in the 
velocity dispersions of cold and intermediate-temperature gas
between the three high surface density models, even though the vertical
gravity varies by a factor 16.  The velocity dispersions of warm gas
vary slightly more, but still by only $50\%$.
The velocity dispersions are somewhat
larger in the outer galaxy model, most noticeably in the cold medium.

\begin{figure}
\plotone{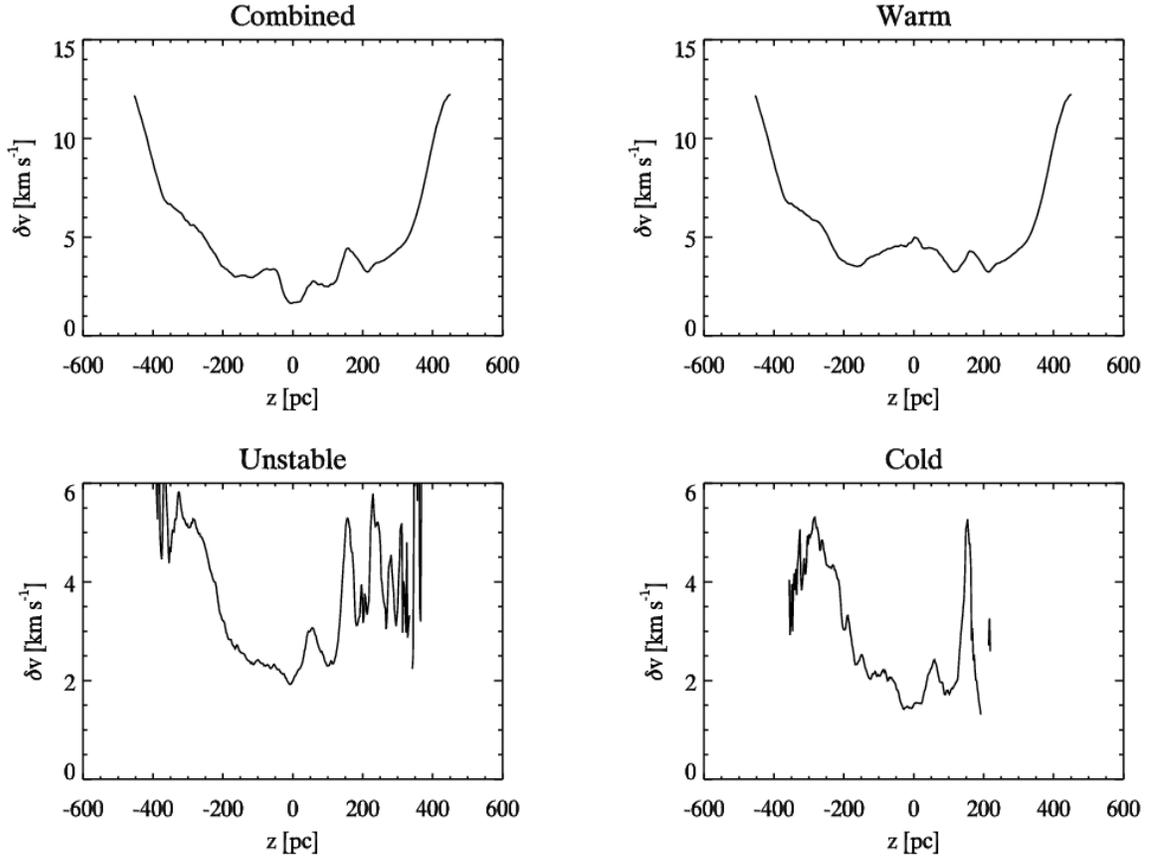}
\caption[Profile of $P_{turb}=\rho\delta v$ for the standard
run]{Vertical profile of mass-weighted turbulent velocity dispersion
$\delta v$ for the standard
model, for all of the gas as well as each of the three thermal
components.  
\label{f5}}
\end{figure}

For the standard model, Figure \ref{f5} shows the late-time ($t=8-10$
orbits) mass-weighted velocity dispersion profile in $z$ for the warm,
unstable, and cold components separately (i.e. $[\sum \rho \delta
v^2/\sum \rho]^{1/2}$ as a function of $z$), as well as for the
combined medium.  For the warm medium, $\delta v$ generally increases
with height, and reaches nearly $12 \ \kms$ near the boundary of the
simulation domain.  This increase with height simply reflects the
(exponential) decrease in density of the gas with height.  The low
inertia of the high-altitude gas allow magnetic stresses to accelerate
it to very high speeds.  Note that the velocity dispersions of cold
and intermediate-temperature gas are asymmetric at high $|z|$.  The
asymmetry in this (and similar) plots is simply due to the relatively
small amounts of cooler gas present at these heights, such that the
turbulent velocity field is not fully sampled statistically (the warm
medium, which samples the velocity field better, yields
fairly symmetric profiles).

We can compare the velocity dispersions of the stratified models to
those of the unstratified models from Paper II.  The total mass
weighted velocity dispersions of the low, middle and high gravity
models are 2.9, 2.7, and 3.0 $\kms$, and the mass-weighted mean
densities\footnote{derived from the profiles of the average density as
  a function of $z$; see \S 3.5.}  are 1.5, 2.7, and 7.0 $\cmt$,
respectively.  The outer-galaxy model has mass-weighted velocity
dispersion 5.5 $\kms$ and mass-weighted mean density 0.24 $\cmt$.
From Paper II we found that the velocity dispersion followed a
relationship $\delta v\propto \bar n^{-0.77}$ with $\delta v = 2.7 \
\kms$ for the same initial magnetic field strength as the present
models, and density $\bar n=1$.  So, based on the non-stratified model
scaling, the velocity dispersion at the corresponding mean densities
would be predicted to vary from 2.0 $\kms$ in the low gravity model,
to 1.3 $\kms$ in the standard model, to 0.6 $\kms$ in the high gravity
model, and to 8.1 $\kms$ in the outer-galaxy model.  The values we
find for the velocity dispersion are within the same general range as
the results from our unstratified models.  However, in detail the
results from our stratified models do not follow the velocity
dispersion scaling with the mean density found in the unstratified
models.  In fact, this is not surprising given the large variation of
mean density with $z$ and the nonlinear relationship between $\delta
v$ and $\bar n$.  At high $z$, as pointed out in Paper II, it is
expected that the velocity dispersion/density relationship will turn
over at velocities comparable to the thermal speed in the warm gas.
Thus, we do not expect velocity dispersions to significantly exceed $8
\ \kms$, and based on Figure \ref{f5} this is the case.

The numbers presented here illustrate the primary difference between
the present simulations and those of Paper II.  In the non-stratified
models of Paper II the three phases of the ISM were well mixed within
the simulation domain, which led to the result that the three phases
of gas were found to have essentially the same turbulent velocities.
In the current stratified simulations, cold gas is found primarily
near the mid-plane, with the low density warm medium dominating the
dynamics at higher latitude.  For all of our models, turbulent
velocities are significantly higher in the low density warm medium
than the high density cold medium.  Because of the differential
stratification of diffuse gas and dense clouds, the large turbulent
velocities in the high-latitude warm gas do not serve to drive
comparably high turbulence levels near the mid-plane, where most of
the cold medium is found.  Even at a given height, the velocity
dispersions in warm and cold gas differ.  Near the midplane ($|z|<100$
pc), the velocity dispersion is a factor two or more lower in the cold
gas than in the warm gas; this is because warm gas mixes more
vertically than the cold gas.

\subsection{Magnetic Fields}

\begin{figure}
\plotone{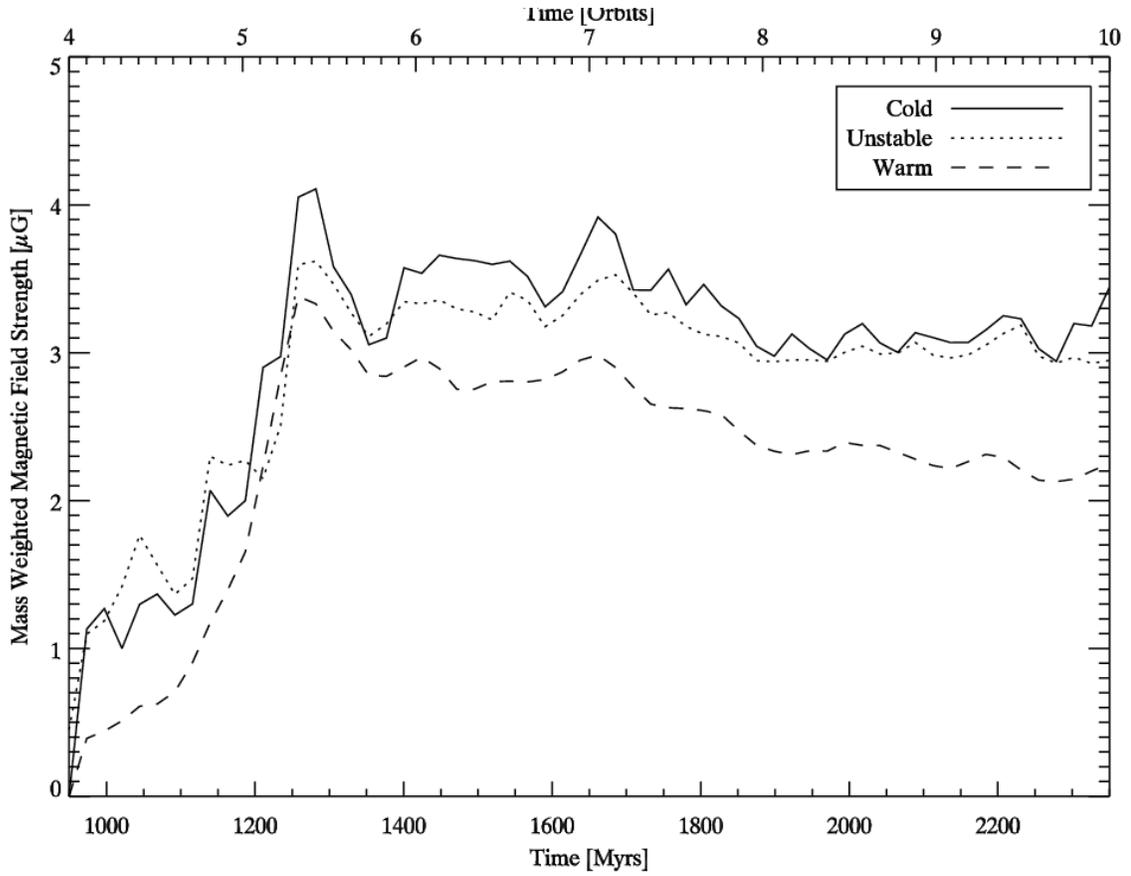}
\caption[Mass-weighted magnetic field strength]{Mass-weighted magnetic
field strength as a function of time for t=4--10 orbits, for the
standard run.  Averaged over $t=8-10$ orbits, the mean field
strengths in the warm, unstable, and cold phases are 2.3, 3.0, and 3.1
$\mu G$.
\label{f6}}
\end{figure}

The initial magnetic field strength in all models is $0.26 \ \mu G$,
and the field is vertical; due to periodic boundary conditions in the
horizontal direction, the value of $\langle B_z\rangle$ is unchanged
over time.  For the standard model, the mass-weighted
magnetic field strength, $B=(B_x^2 + B_y^2+B_z^2)^{1/2}$ is plotted in
Figure \ref{f6}, as a function of time.  Over the course of the 
simulation, the  MRI amplifies the
initial field by an order of magnitude. The saturated state field
strength is typically $3 \ \mu G$, slightly higher in the cold medium,
and slightly lower in the warm medium.  Averaged over orbits 8--10, the
mean field strengths in the warm, unstable, and cold phases are 2.3,
3.0, and 3.1 $\mu G$.  In the low gravity model the field strength
values are 2.8, 3.0, and 3.1 $\mu G$, and in the high gravity model we
find slightly larger means of 2.8, 3.5, and 3.7 $\mu G$, respectively.
Magnetic field strengths are slightly lower in the outer galaxy model, where
we find 2.1, 2.5, and 2.7 $\mu G$, respectively.  Overall, there is much less
variation of magnetic field strength with mean density or phase than the 
variations we find in velocity dispersions.  This result is consistent with
our findings from our unstratified models, where we also found very little 
variation in the saturated-state values of $B^2$.  Thus, the
saturated-state value of $B^2$ appears to be controlled by the
value of the midplane thermal pressure -- which, due to the heating
and cooling processes involved, is similar for all models.

\begin{figure}
\plotone{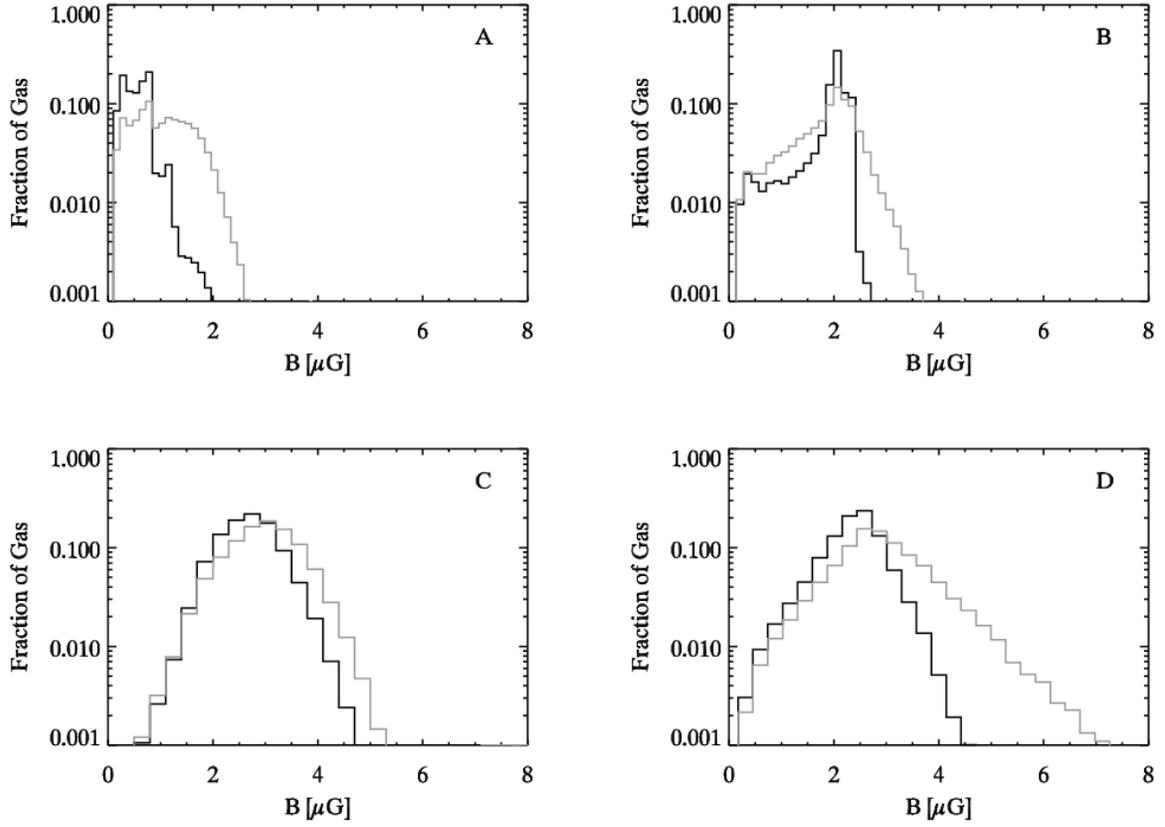}
\caption[Magnetic field PDFs for the standard run]{Volume weighted
(dark line) and mass-weighted (grey line) magnetic field PDFs for the
standard run, at t=4.5, 5.0, 7.5 and 10.0 orbits (panels A, B, C, and
D, respectively).
\label{f7}}
\end{figure}

In Figure \ref{f7} we show the PDF of the magnetic field strength at
$t=4.5$, 5.0, 7.5 and 10.0 orbits, for our standard model.  The 
breadth of the PDFs grow with time until 7 orbits.  After that point, a tail 
in the mass-weighted PDF at high $B$-values (up to $8 \ \mu$G) develops, but 
the volume-weighted PDF as well as the mass-weighted mean $B^2$ remains 
nearly constant.  

\subsection{Distributions of Density, Temperature, and Pressure}

\begin{figure}
\plotone{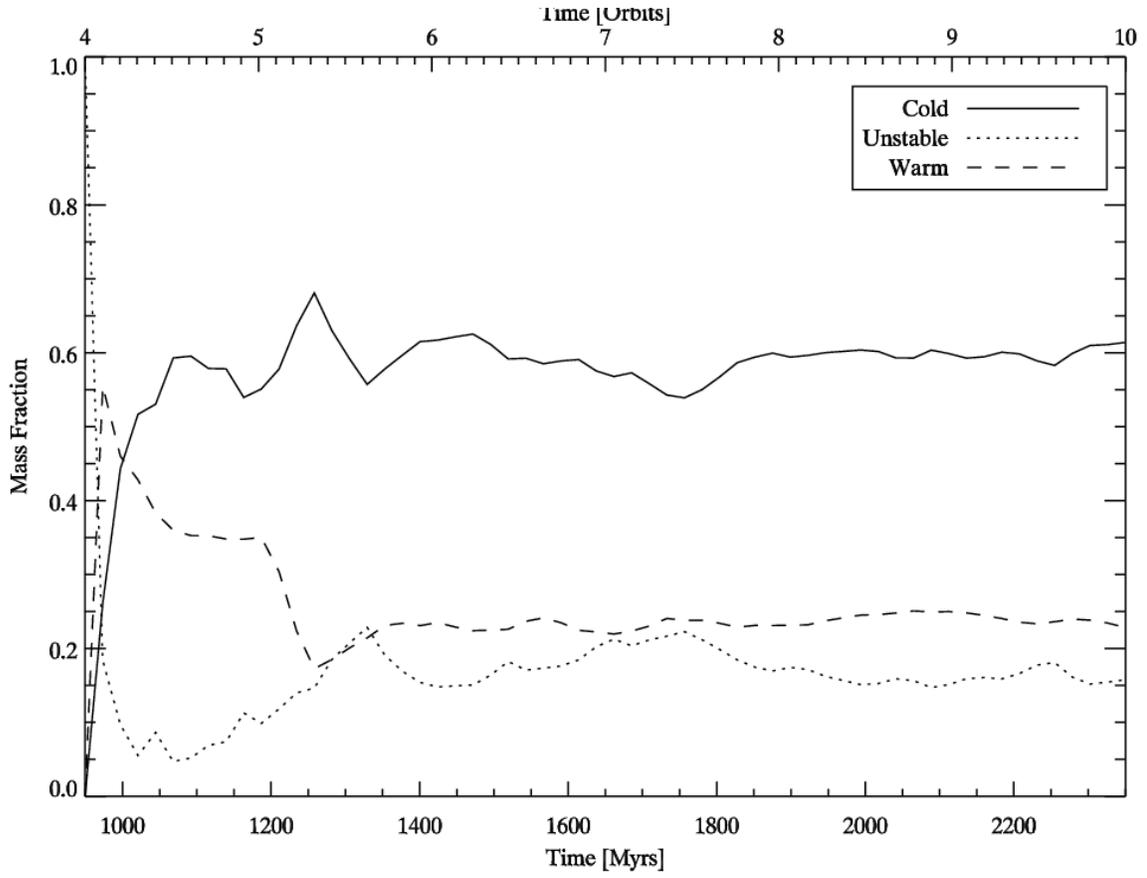}
\caption[Mass fractions of the three phases of gas]{Mass fractions of
the three phases of gas for the standard run.  Averaged over orbits
8--10, the warm, unstable, and cold phases contain 24\%, 16\% and 60\%
of the mass, respectively.
\label{f8}}
\end{figure}

In Figure \ref{f8} we plot the mass fractions of the three phases in
the standard model as a function of time, from $t = 4-10$ orbits.  By
mass the warm medium is 24\% of the total, with the unstable and cold
media providing 16\% and 60\% of the mass, averaged over orbits 8--10.
By volume the warm phase occupies 92\%, with the unstable phase about
6\% and the cold phase about 2\% (this result, of course, depends on
the chosen size of the computational box).  For comparison, in the
high and low gravity models, the mass fractions of warm, unstable, and
cold gas are (16,11,73)\% and (22,18,60)\%, respectively. Thus, the
proportions of mass in different phases appears fairly insensitive to
${\bf g}_{ext}$.  We can contrast this with the results of the
non-turbulent comparison models.  For those models, the cold mass
fraction was 81\%, 86\%, and 92\% for the low, medium, and high
gravity models, with the warm fraction making up the balances of 19\%,
14\%, and 8\%, respectively.\footnote{Note that while the high- and
medium-gravity non-turbulent models have warm fractions generally
consistent with 
the prediction and scaling of equation (\ref{Sigw}), the low-gravity case has a
warm fraction much lower than this equation would predict.  This is
simply because the distance of 450 pc from the midplane to the top of
the computational box is less than the warm medium scale height of 590
pc, so that the volume available for warm gas is limited.}  Thus,
turbulence considerably lowers the fraction of gas found in the cold
regime for these high-surface-density models.

For the outer galaxy model model at late times, we find mass fractions of
(64,20,16)\% for the warm, intermediate, and cold phases, respectively.  All
of our models are initialized as isothermal and are therefore out of in
thermal equilibruim. When cooling is turned on at $t=4$ orbits, depending on
the density and temperature, local regions in the disk will either heat or
cool towards equilibrium.  In the case of the outer galaxy model, unlike the
standard model, the low mean surface density and large scale height of the
initial distribution yields a medium which exists entirely in the warm regime
after cooling is turned on. The initial central pressure is $P_0/k= 650 {\rm \ 
  K} \ \cmt$, and this rises quickly to approximately $P_0/k= 1800 {\rm \ K} \ 
\cmt$ shortly after cooling is enabled, and then decreases somewhat.  Later in
the simulation, turbulence from the MRI forces some of the gas to condense
into the cold phase.  For this low surface density model, however, the total
mass fraction of cold gas remains low; the warm medium dominates.

\begin{figure}
\plotone{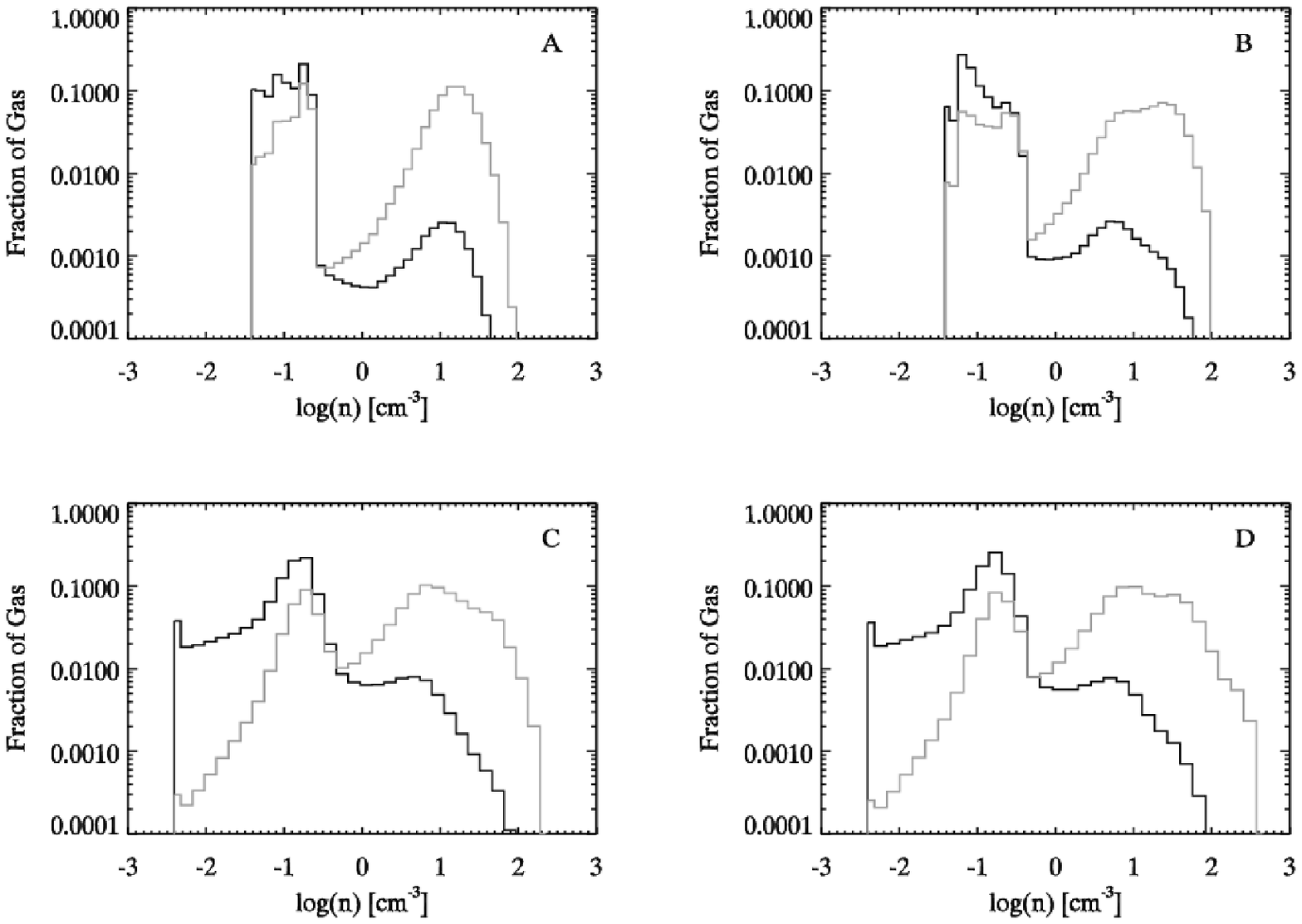}
\caption[Density PDFs for the standard run]{Volume weighted (dark
line) and mass-weighted (grey line) density PDFs for the standard
model, at $t=4.5$, 5.0, 7.5 and 10.0 orbits (panels A, B, C, and D,
respectively).
\label{f9}}
\end{figure}

In Figure \ref{f9} we plot the density PDF at $t=4.5$, 5.0, 7.5, and 10.0
orbits for the standard model. These PDFs indicate the presence
of two distinct phases of gas, as there are peaks in the volume-weighted
PDF near $n=0.15\ \cmt$ and a broader peak in the mass-weighted PDF centered on
$n=15 \ \cmt$.  The minimum density reaches
the artificial density floor of 0.004 $\cmt$, while the maximum density can
extend upwards of 400 $\cmt$. Overall these PDFs are similar to those of the
non-stratified models of Paper II.  The main difference in the results of the
present models
is the tail extending to low densities (due to gravitationally-imposed
stratification).  The density PDFs for the high- and low-gravity
models are similar.
Taking into account the large differences in the total 
mass and volume fractions
between the outer galaxy model and the standard model, the density PDFs of the
outer galaxy model are also 
qualitatively similar.  The peaks in the distributions
corresponding to the warm and cold phases lie in approximately the same
location.  However, the distribution of cold gas falls off more quickly at
higher densities compared to the standard model case.

\begin{figure}
\plotone{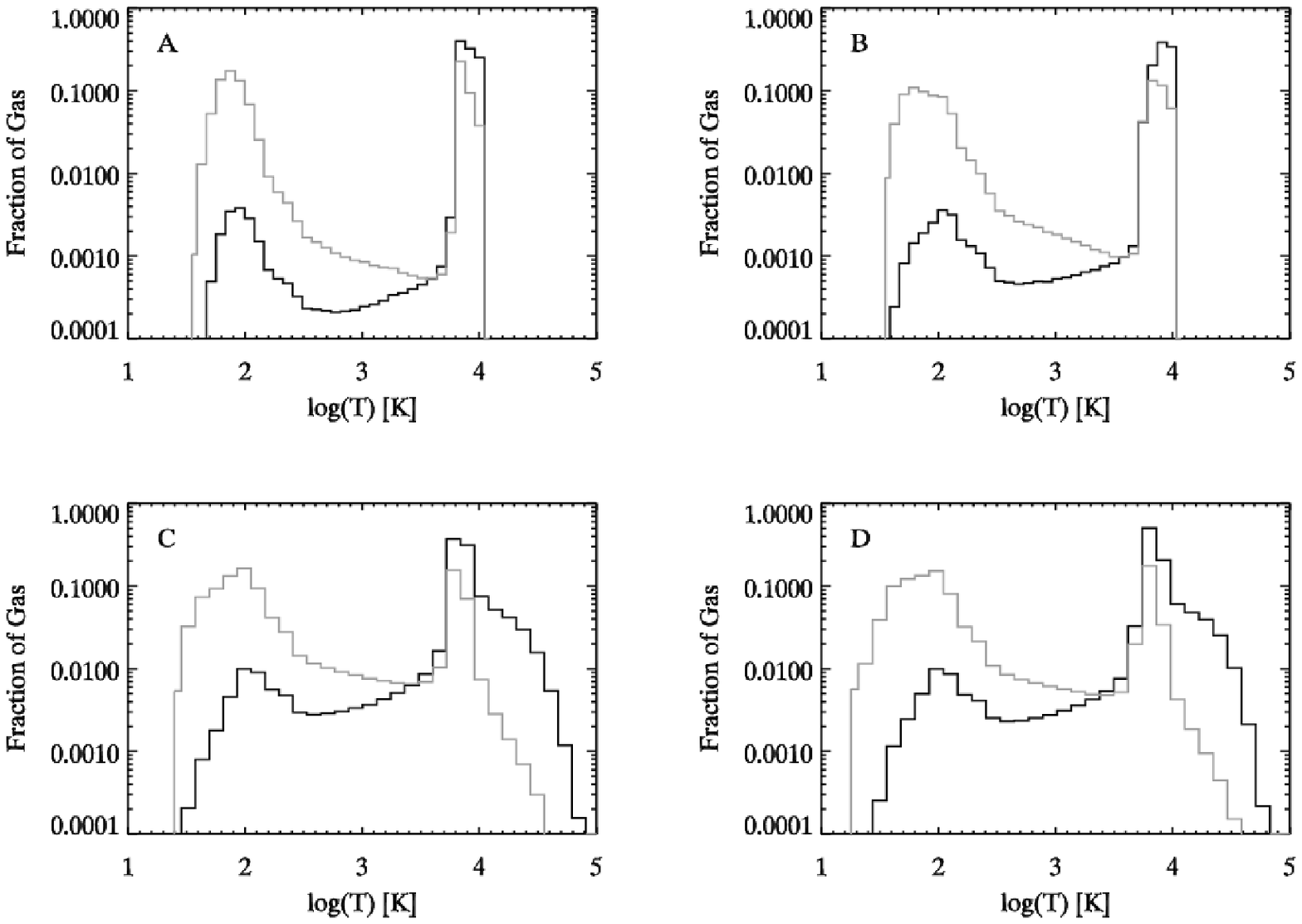}
\caption[Temperature PDFs for the standard run]{Volume weighted (dark
line) and mass-weighted (grey line) temperature PDFs for the standard
model, at $t=4.5$, 5.0, 7.5 and 10.0 orbits (panels A, B, C, and D,
respectively).
\label{f10}}
\end{figure}

In Figure \ref{f10} we plot the temperature PDFs at the same times
as those presented in Figure \ref{f9}.  The high temperature peak is
fairly well defined.  This feature is broadened later in the
simulation, with a small fraction of the gas existing at 
temperatures higher by  an order of magnitude  compared to the PDFs
from Paper II.  In part, this is because the equilibrium temperature of 
high-altitude, low-pressure gas is higher than that of the warm gas near
the midplane, for our adopted cooling curve. For the outer galaxy model
the temperature PDFs are generally similar to those of the standard
model.  The peak in the cold medium is found in the same location, but
towards cooler temperatures the distribution is somewhat
truncated compared to that of the standard model.

\begin{figure}
\plotone{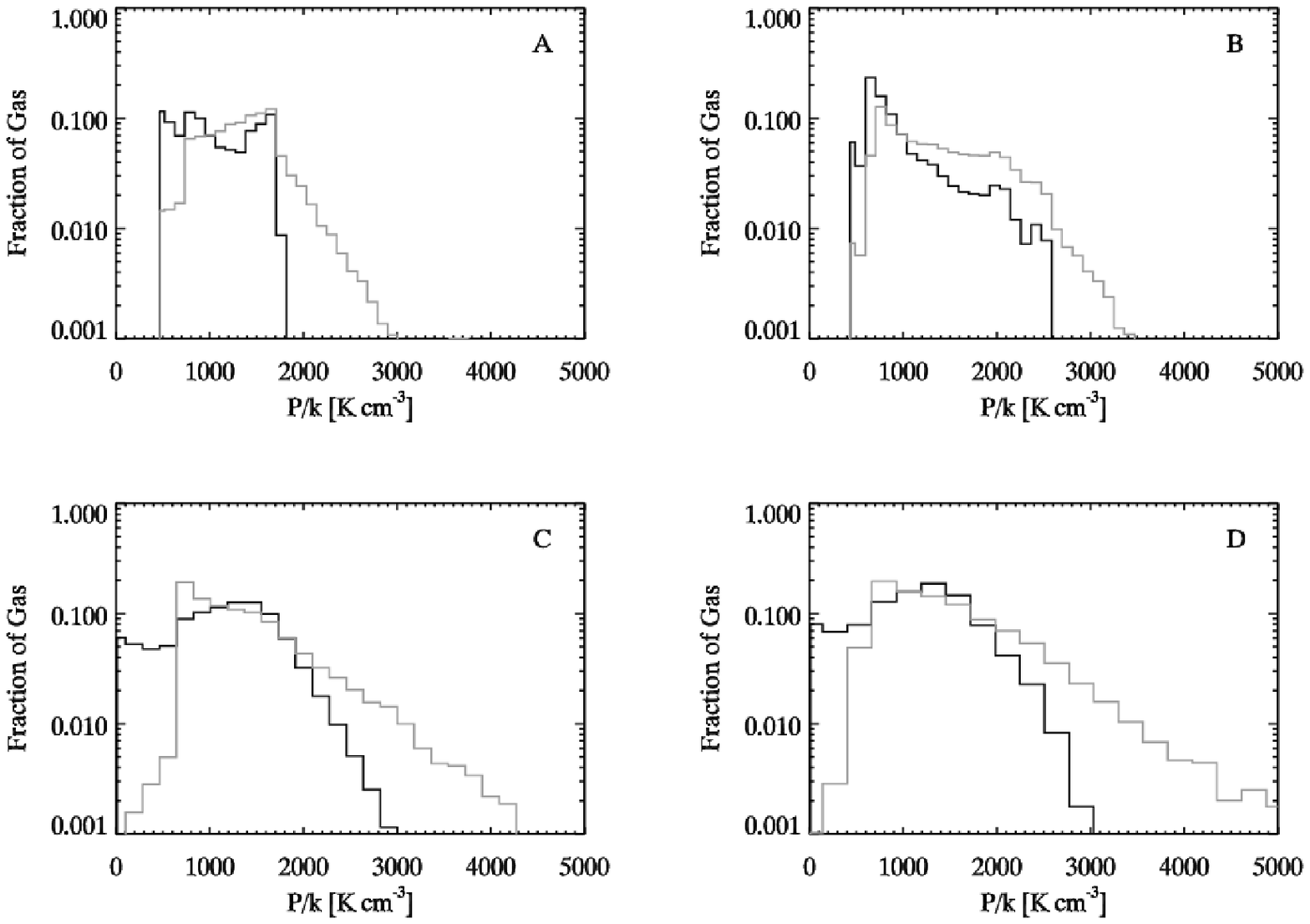}
\caption[Pressure PDFs for the standard run]{Volume weighted (dark
line) and mass-weighted (grey line) pressure PDFs for the standard
model, at t=4.5, 5.0, 7.5 and 10.0 orbits (panels A, B, C, and D,
respectively).
\label{f11}}
\end{figure}

\begin{figure}
\plotone{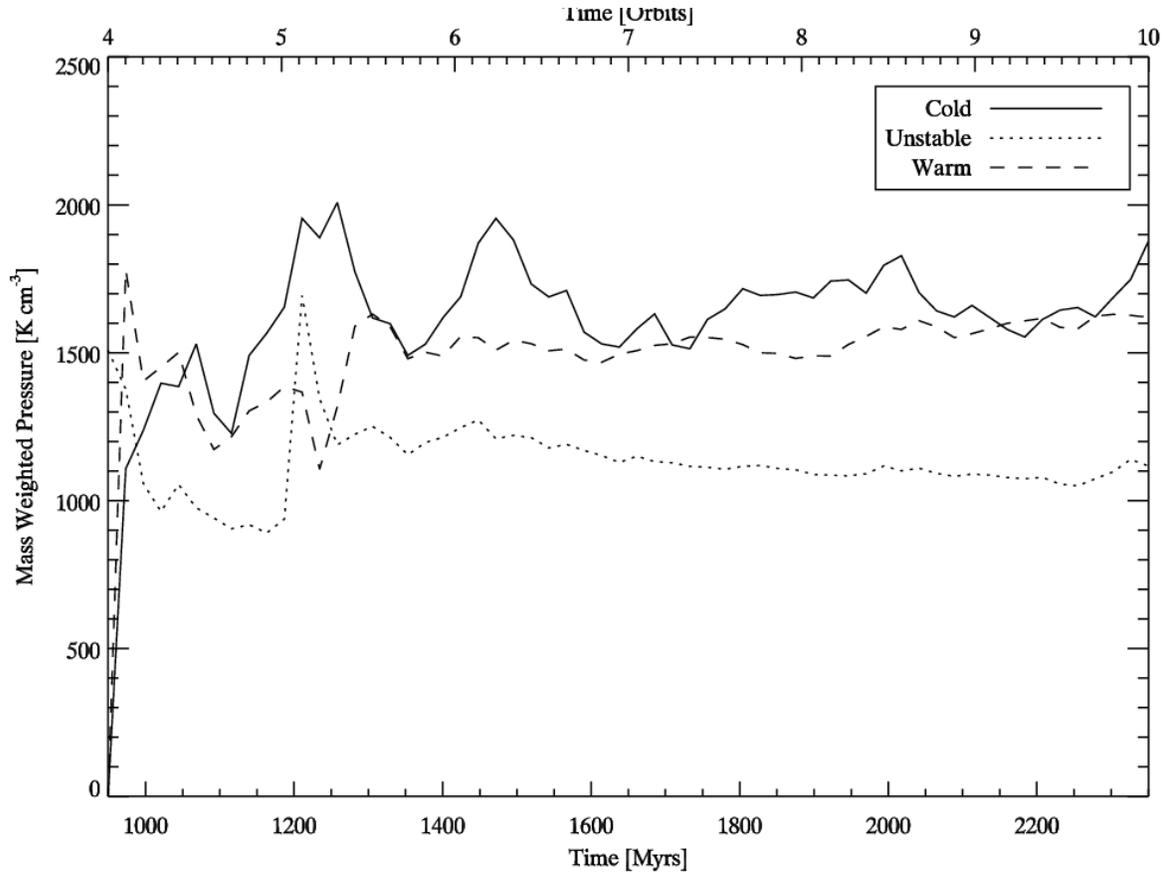}
\caption[Mass-weighted pressure as a function of time for the standard
run]{Mass-weighted pressure as a function of time for the standard
run. Averaged over orbits 8--10, the mean pressures are $P/k=1600,
1100,$ and $1700\ \rm{K \ cm^{-3}}$ for the warm, unstable and cold
phases.
\label{f12}}
\end{figure}

The pressure PDFs for the standard model 
are shown in Figure \ref{f11}, again at $t=4.5$, 5.0, 7.5,
and 10.0 orbits.  The pressure PDFs extend to very low values in the saturated
state, with $\sim 20\%$ of the gas by volume (but only a few percent by mass)
below $P/k=500\ \rm{K \ cm^{-3}}$.  The maximum pressure is approximately
$P/k=8000\ \rm{K \ cm^{-3}}$, but only a few percent of the mass is at
pressures above $P/k=3000\ \rm{K \ cm^{-3}}$.  The maximum pressure for which
a warm medium can exist in thermal equilibrium (for our cooling function) is
$P_{max}/k=3100\ \rm{K \ cm^{-3}}$; this is where the volume weighted PDF cuts
off, implying negligible gas is above this pressure.  Also, the mass-weighted
fraction of gas drops sharply below $P_{min}/k=800\ \rm{K \ cm^{-3}}$, the
lowest pressure for which a cold medium can exist in thermal equilibrium. The
mass-weighted mean pressure in the warm, cold, and unstable phases for
the standard model is shown in
Figure \ref{f12}.  The mass-weighted mean pressures, averaged over orbits
8--10, for the warm, unstable, and cold phases are $P/k=1600, 1100,$ and
$1700\ \rm{K \ cm^{-3}}$, respectively.  So, although the overall pressure
distribution is quite broad, the mean pressures in the warm and cold phases
are approximately equal.  Our results are similar for the other two
high-surface-density runs.  In the high gravity run the mean warm, unstable,
and cold pressures are  $P/k=1500, 1200,$ and $2200 \ \rm{K \ cm^{-3}}$,
respectively; and in the low gravity case the corresponding mean values for
each phase are $P/k=1500, 1100,$ and $1400 \ \rm{K \ cm^{-3}}$, respectively.
Interestingly, the mean pressures in ``transitional'' gas (i.e. in the
thermally unstable temperature range) are lower than pressures in both the
warm and cold phases, and similar to the transition pressure for
the non-turbulent case.  

In the outer galaxy model the mass-weighted pressure PDF does not
extend above $P/k=3000 \ \rm{K \ cm^{-3}}$, unlike the results shown
in Figure \ref{f11} for the standard model.  The volume weighted
PDFs are very similar in the standard and outer galaxy models,
however, and the mass-weighted PDF of the outer galaxy model in fact
follows its volume weighted PDF closely at pressures higher than
$P/k=1000$.  The mean mass-weighted warm, unstable, and cold pressures
for the outer galaxy model are $P/k=1500, 1200,$ and $1100 \ \rm{K \
cm^{-3}}$, averaged over the final two orbits of the simulation.
Thus, for this model in which the warm phase dominates the total mass,
the mean warm-phase pressure slightly exceeds that of the cold phase.

\begin{figure}
\plotone{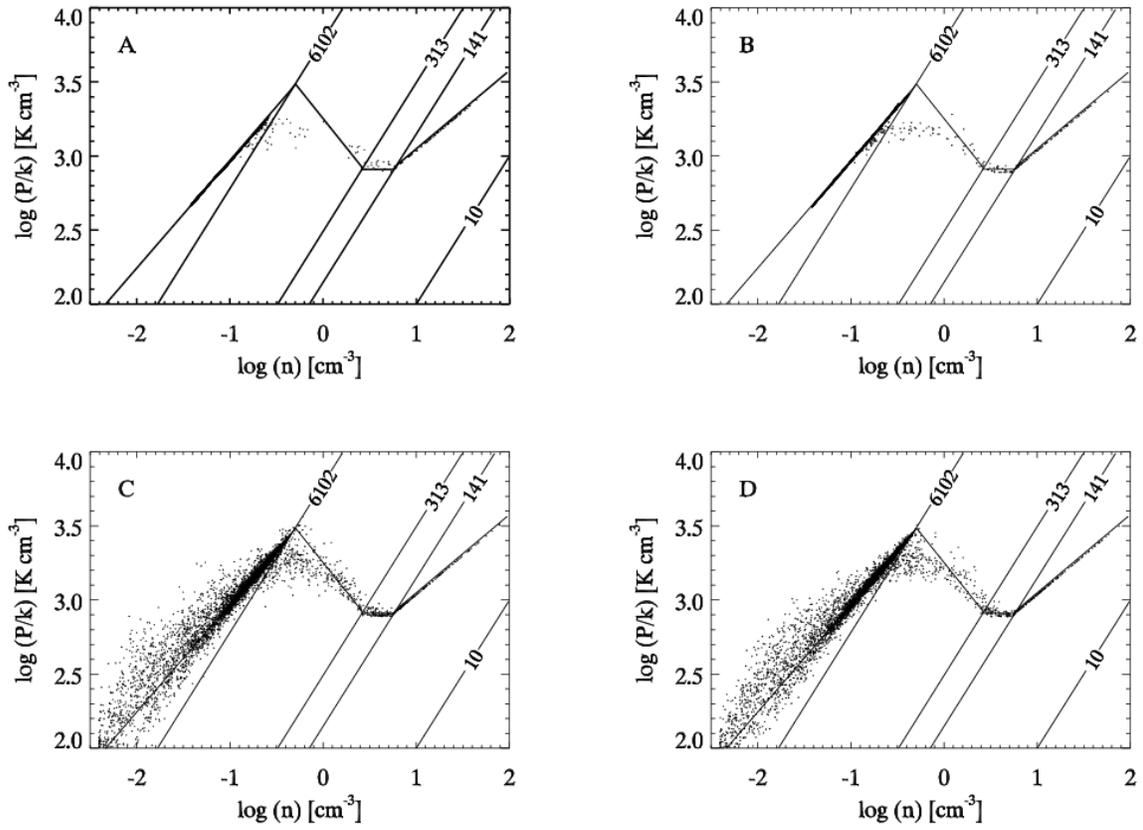}
\caption[Scatter plots of density vs pressure for the standard
model]{Scatter plots of density vs pressure for the standard model, at
$t=4.5, 5.0, 7.5$ and 9.9 orbits (panels A, B, C, and D, respectively).
\label{f13}}
\end{figure}

Scatter plots of density vs pressure for the standard run are shown in
Figure \ref{f13}.  The solid line is the equilibrium cooling curve,
and contours of constant temperature are plotted at the transitions
between different temperature regimes in the cooling function.  
Gas at low temperatures
cools relatively quickly, and is found to be very close to the thermal
equilibrium curve.  For the first orbit after the cooling function is
enabled, during which turbulent amplitudes are relatively low, most of
the gas is in thermal equilibrium in both the warm and cold
phases. (Note that the transition from isothermal to this two-phase
state occurs very rapidly, due to thermal instability, after cooling
is enabled.)  Later on, turbulence drives significant fractions of gas
out of thermal equilibrium. Gas at higher temperatures takes longer to
cool and is typically out of equilibrium, although is still roughly
follows the shape of the equilibrium curve in the $P-\rho$ plane.
Cold gas, with its short cooling time, always closely follows that
thermal equilibrium locus.  Scatter plots for our other models,
including the outer galaxy case, do not show any significant
differences from those of the standard model.

For all of our models, in the turbulent saturated state the
mass-weighted mean value of pressure lies in the range for which two
stable phases are possible.  Quantitatively, this mean pressure is
typically near the geometric mean of $P_{min, cold}$ and $P_{max,
warm}$.  Compared to the non-turbulent situation, for which the mean
pressure in the warm medium is lower (by a factor $\sim 2$), an
increase of pressure at fixed temperature implies an increase in the
mean warm density.  Coupled with an increase in the warm medium's
scale height (due primarily to magnetic support), this tends to
increase the fraction of gas in the warm phase compared to the non-turbulent
case.

\subsection{Stratification of Density and Pressure: What Supports Gas 
Vertically?}

To address the issue of vertical support of the ISM, we first
summarize results from our non-turbulent models that do not include
MRI driven turbulence.  Once the gas has settled, we can evaluate mass
distributions (and total mass fractions; see \S 3.4) of each
phase. The profiles of both the cold medium and warm medium are close
to truncated Gaussians, with a narrow transition layer in the
thermally unstable phase.  While the cold mass fraction is higher when
gravity increases, the cold disk thickness is smaller; the cold scale
height decreases from 40 to 20 to 10 pc from the low- to medium- to
high- gravity case.  The transition between phases occurs at $P/k =
1100-1200 \ \rm{K \ cm^{-3}}$, i.e. close to the minimum pressure for
which cold and warm gas can coexist.  The warm disk has a scale height
of 460, 240, and 120 pc in the low, mid, and high
gravity models, respectively.\footnote{Note that these scale heights
  are computed by fitting to the density profiles; fits to the
  pressure profiles yield slightly larger scale heights of 
 580, 290, and 145 pc, respectively.}  
These successive factors of two in the
scale heights of cold and warm components are expected based on the
successive factors of 4 in $\tilde g$ from low to moderate to high
gravity, since for an isothermal pressure-supported disk,
$H=c_s/\sqrt{ \tilde{g}}$, where ${\bf g}_{ext}\equiv-\tilde{g}z$.  We
can compare the scale heights of these non-turbulent models to those
which include MRI driven turbulence.

\begin{figure}
\plotone{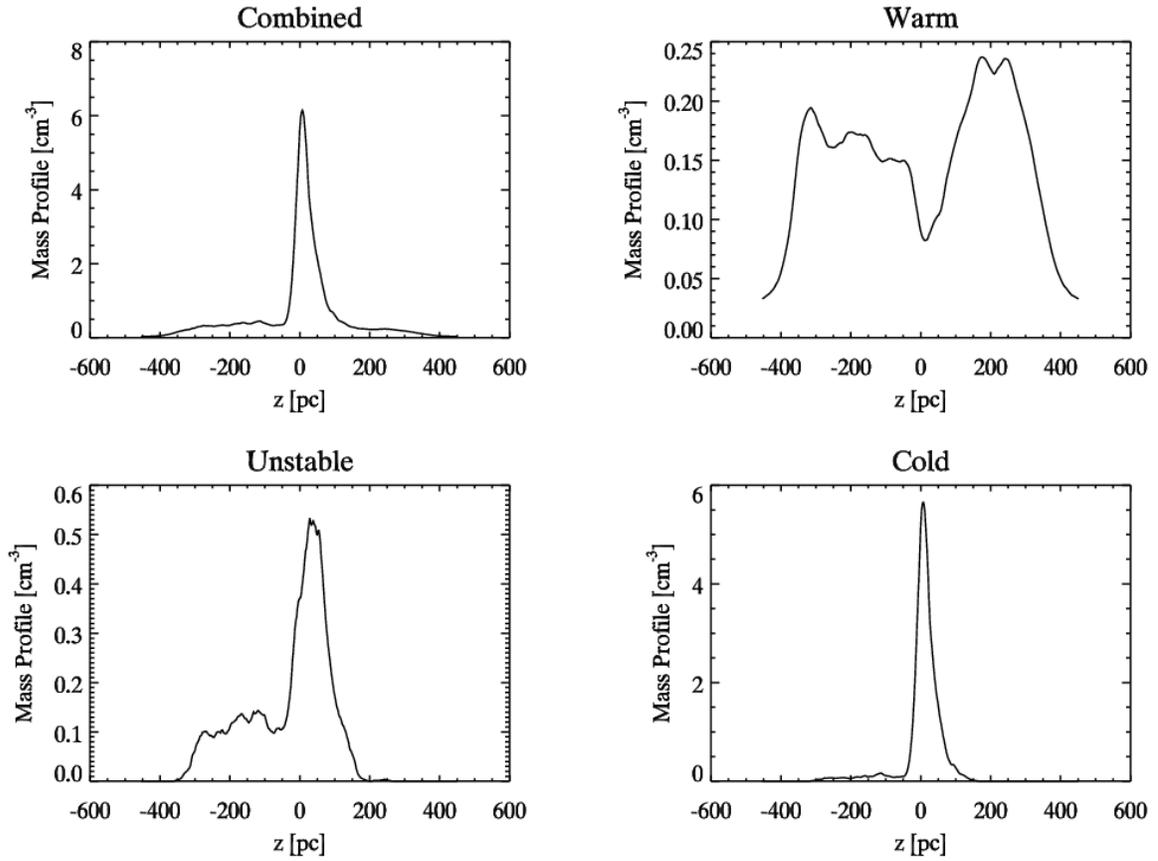}
\caption{Mass profiles for the standard model.  The warm and unstable phases of
gas have obviously non-Gaussian profiles; for the cold phase we find
$H \sim 20$ pc.
\label{f14}}
\end{figure}

Mass profiles for the three phases of gas as well as the total are
shown in Figure \ref{f14} for our standard turbulent model.  These
profiles are computed by integrating the density in each component in
$x$ and $y$, as a function of $z$, dividing by the total number of
zones in the $x$ and $y$ directions, and averaging over the last two
orbits of the simulation.  In the turbulent models, the scale height
of the cold gas is 50, 20, and 12 pc from low to high gravity.
Turbulence from the MRI therefore only slightly increases the scale
height of the cold medium.
The profile of the unstable gas is also non-Gaussian, but we
estimate a scale height of the centrally peaked gas to be 90, 44, and
30 pc from the low to high gravity models.  These values are roughly
twice as large as the cold layer thickness.  There is significant
unstable gas at high latitude.  For the low gravity case the warm
medium is more or less evenly distributed vertically through the box.  
In both the standard gravity case and the high gravity case, the 
warm gas is very far from Gaussian,  and has a local minimum at the
midplane.  

\begin{figure}
\plotone{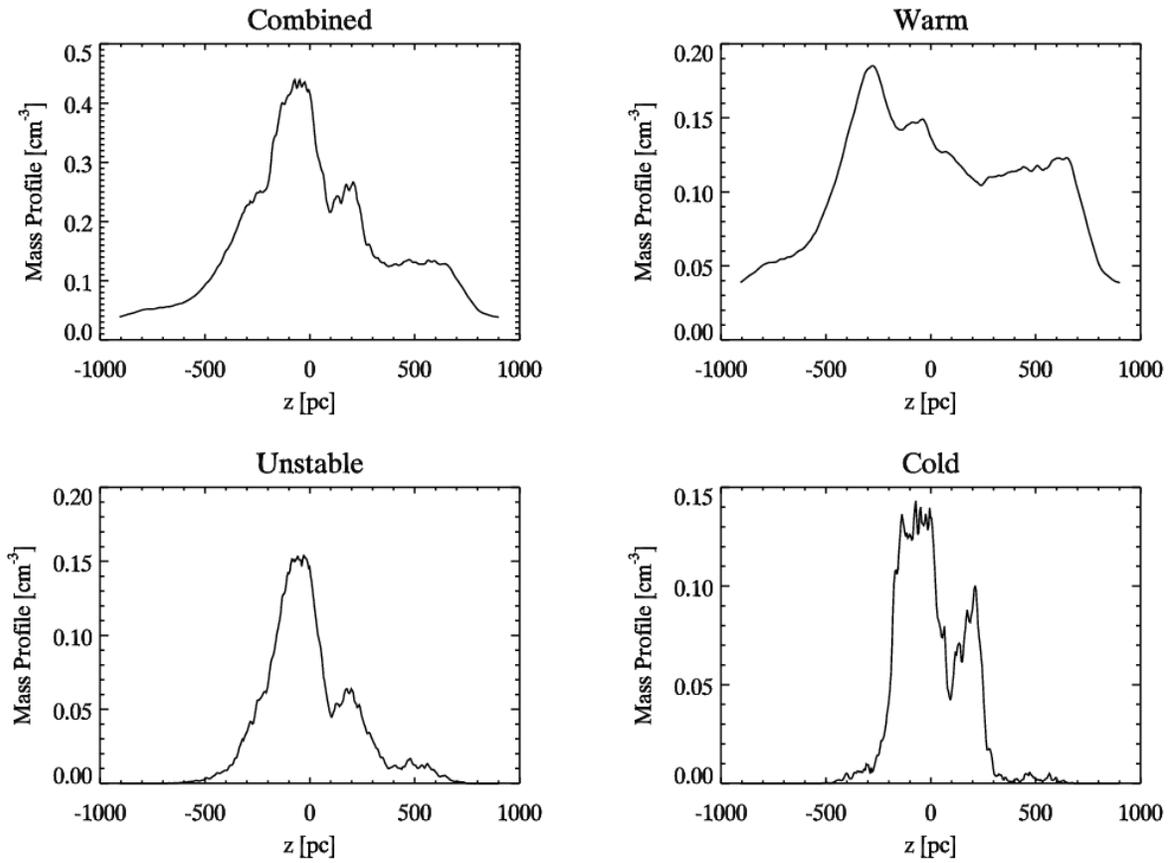}
\caption{Mass profile for the outer galaxy model.  Note the larger
  simulation domain compared to Fig. \ref{f14}.  The cold gas
  distribution is significantly more vertically extended compared to
  our higher surface density models.
\label{f15}}
\end{figure}

We also show mass profiles for our outer galaxy model in Figure
\ref{f15}.  The total mass distribution is broadened significantly
compared to the standard model, and the mean densities are
approximately a factor ten smaller.  The cold mass profile peaks
slightly below 0.15 $\cmt$, compared to 6 $\cmt$ for the standard
model, indicating that the filling fraction of cold gas in the outer
galaxy model near the disk mid-plane is significantly smaller than in
the standard model.  In the outer galaxy model 
the distribution of the cold
medium also extends much farther above and below the mid-plane, dropping
sharply at $\sim 200$ pc.  Extended tails at large $|z|$ are also
present for both cold and unstable gas.  Gaussian fits to the cold and unstable
phases yield scale heights of $\sim 150$ pc each.  In each
case, we fit to the entire mass distribution, neglecting the
local minima at around 100 pc.  For the cold component
our fit overestimates the true distribution at higher latitudes, and
underestimates it closer to the mid-plane.  We emphasize, again, that these
numbers should only be considered to be rough estimates.

\begin{figure}
\plotone{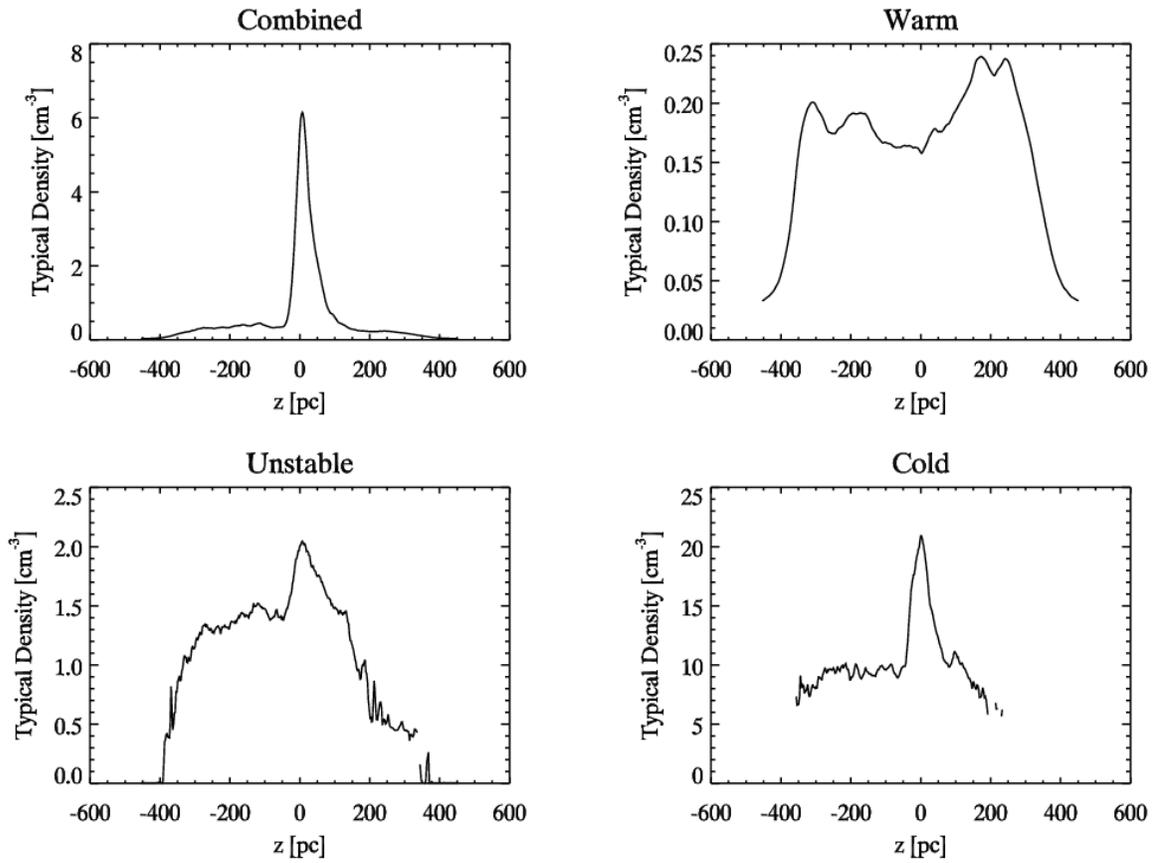}
\caption[Profiles of the typical density, for each phase in the
standard run]{Profiles of the typical density at each height for each
gas phase in the standard model.  The typical density of the cold medium
is approximately 10 $\cmt$, increasing sharply near the mid-plane to
more than 20 $\cmt$.
\label{f16}}
\end{figure}

In Figure \ref{f16} we plot for the standard model the profile of
typical density as a function of height for all the gas, as well as
profiles of typical density in the warm, unstable, and cold components
separately.  The component typical densities are just the mean values
in each phase at any $z$.  The typical density in the warm medium is
approximately $0.2-0.25 \ \cmt$, decreasing at $z>300$ pc.  In the
unstable phase the typical density is around $1.5-2.0 \ \cmt$, increasing
somewhat towards the mid-plane and decreasing at higher $z$.  For the
cold medium the average density reaches as high as about 20 in the
mid-plane, sharply decreasing to around 10 at higher latitudes. This
is near the minimum possible density at which cold gas can be in
thermal equilibrium (at $P_{min, cold}$).  The typical density in the warm
and cold phases differ by two orders of magnitude.  In the outer
galaxy model the typical density of the cold medium is fairly uniform
at around 8--10 $\cmt$, and does not show a strong central peak.

We next turn to what, physically, is responsible for these vertical
mass profiles.  By averaging the $z$ component of the momentum
equation in horizontal planes, and making use of shearing periodic
boundary conditions, we obtain
\begin{equation}
\frac{\partial}{\partial t}\left\langle \rho v_z \right\rangle =
-\frac{\partial}{\partial z}\left\langle \rho v^2_z\right\rangle -
\frac{\partial}{\partial z}\left\langle P\right\rangle -
\frac{\partial}{\partial z}\left\langle \frac{B^2}{8\pi}\right\rangle
+ \frac{\partial}{\partial z}\left\langle
\frac{B^2_z}{4\pi}\right\rangle -\left\langle |{ g}_{ext}| \rho
\right\rangle.
\label{eq1}
\end{equation}
Thus, kinetic, thermal, and magnetic stresses can all contribute to
vertical support of the disk.  Note that the term $\partial_z\langle
B_z^2/(4\pi)\rangle$ arises due to the vertical magnetic tension force; where
the horizontally-averaged magnitude of $B_z^2$ increases {\it upward}
(magnetic ``hammock'' geometry, in which field lines are more
horizontal near the midplane and more vertical at large $|z|$), a net
upward tension force is exerted on the medium.

\begin{figure}
\plotone{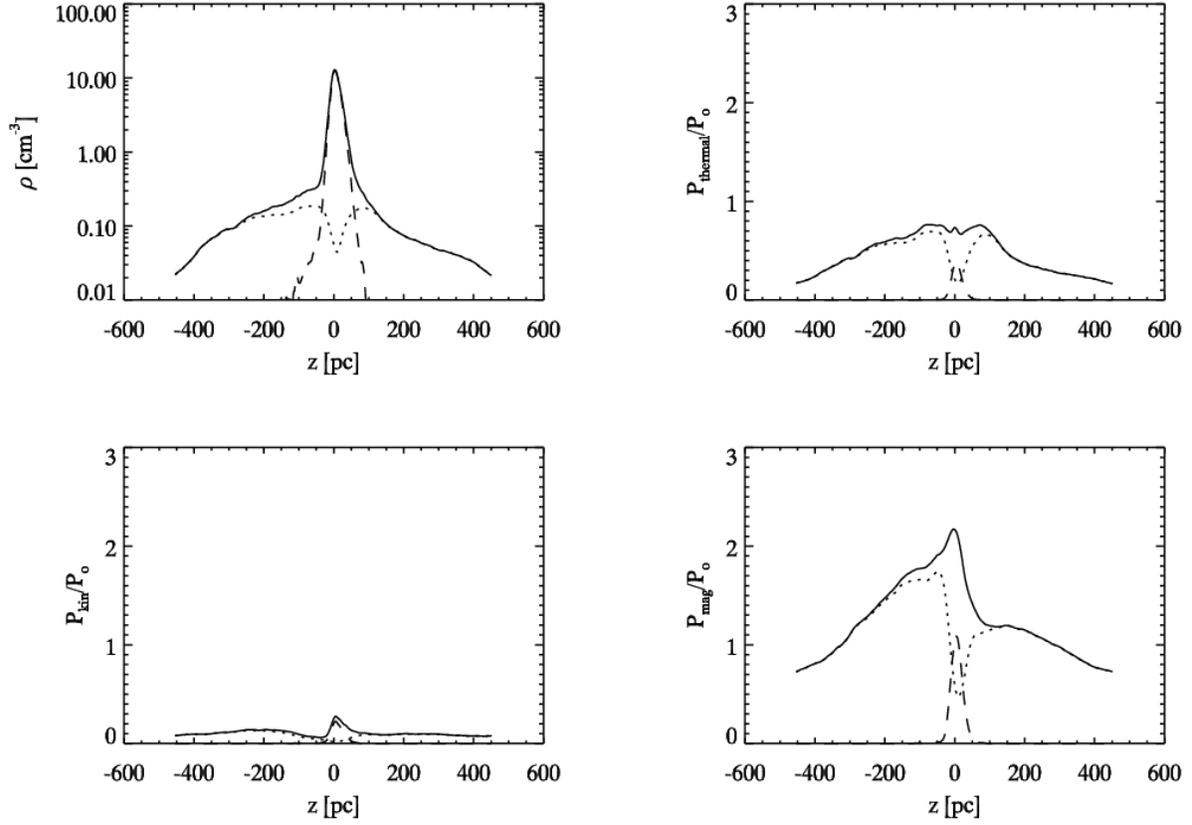}
\caption[Density and pressure contributions from warm and cold phases
for the high gravity run]{Density and pressure contributions from warm
and cold phases for the high gravity model. The warm phase corresponds
to the dotted line, while the cold phase is shown as the dashed line.
The thick solid line shows the total for the warm, unstable, and cold
components combined.
\label{f17}}
\end{figure}

The contributions to thermal, kinetic (i.e. $\rho v_z^2$), and
magnetic pressures for the high gravity model (averaged over $t=8-10$
orbits) are plotted in Figure
\ref{f17} for the warm and cold phases, along with the density
profile. At each height, the contribution from each phase consists of
the sum over zones in that phase, divided by the total number of
zones.  The warm medium dominates the density profile at high $|z|$,
while the cold medium dominates near the midplane.  For both the warm
and cold components, the magnetic pressure is the largest of the three
pressures, followed by thermal, and then kinetic pressures.
At $z=0$, the ratio of mean thermal to mean magnetic pressure is
$\beta=0.3$; this drops to $\beta=0.2$ at $|z|=z_{max}$.

\begin{figure}
\plotone{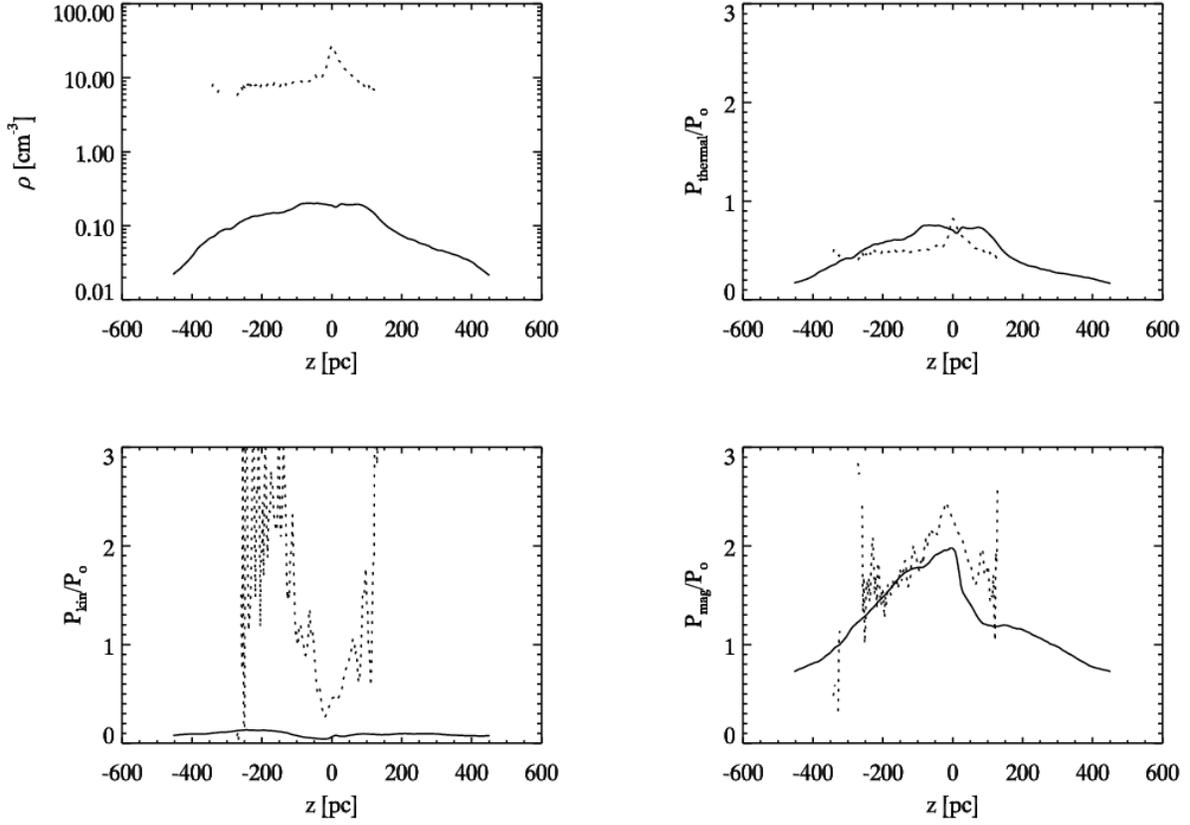}
\caption[Typical density and pressure for the high gravity
model]{Typical density and pressure for the high gravity model.  The
warm phase corresponds to the solid line, while the cold phase is
shown as the dotted line.
\label{f18}}
\end{figure}

In Figure \ref{f18}, we show the vertical profiles of the same
quantities shown in Figure \ref{f17} averaged separately over each
phase.  Similar to the standard gravity run (see Fig. \ref{f16}),
the typical density of the cold medium is around $10-20 \ \cmt$,
increasing significantly towards the mid-plane.  The typical density
of the warm medium is around $0.1-0.2 \ \cmt$ within a few 100 pc of the
midplane, decreasing at higher $|z|$. The kinetic pressure of the cold
medium can be very large, but this is only at high $z$ where little
cold gas is actually present.  The kinetic pressure is large here as the
cold medium is driven to approximately the same velocity as the warm
medium, but has a much higher density than the warm medium.  The
kinetic pressure in the warm gas varies less in $z$ than any other
pressure, but is only 20\% of the mid-plane thermal value.  Away from
the midplane, the kinetic pressure of cold gas exceeds its
thermal pressure.  The thermal pressures of
the warm and cold phases are approximately equal near the
mid-plane.  (Note that the ``typical'' midplane pressures may be
slightly lower than the average values reported in \S 3.4, because the
latter is based on a mass-weighted mean.)
The thermal pressure decreases quickly for the cold medium
away from the mid-plane, while tapering off more slowly in the warm
medium.  The typical magnetic pressures in the warm and cold
phases have similar magnitudes and profiles.  At the midplane, the cold medium has $\beta=0.3$, and the
warm medium has $\beta=0.4$.

Generally, the same trends and behaviors are seen in the standard
gravity run as for the high-gravity case.  For the low-gravity and
outer-galaxy models, the thermal and magnetic pressures show much less
central concentrations.  However, as is true for the high-gravity
case, all models are centrally magnetically dominated: for all
components and all models, the midplane values of $\beta$ are in the
range 0.3-0.6.

\begin{figure}
\plotone{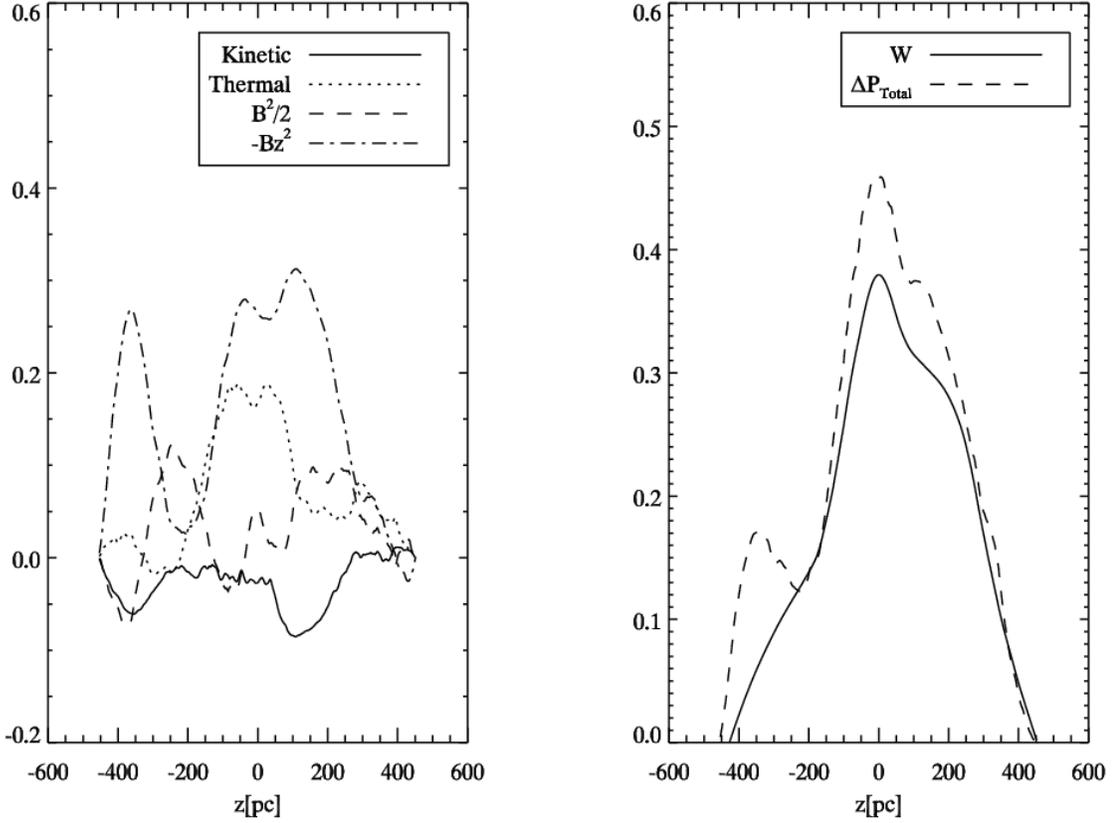}
\caption{On the left we show for the low gravity model the
  contributions arising from the four pressure terms in equation
  (\ref{eq2}), $\rho v_z^2$ , $P_{th}$, $(B_x^2 + B_y^2 + B_z^2)/8\pi$,
  $-B_z^2/4\pi$ (respectively denoted as Kinetic, Thermal, $B^2$, and
  $-B_z^2$.)  For each term, $\Delta P(z)\equiv P(z) -P(z_{max})$ is
  plotted.  On the right we plot $\Delta P_{tot}(z)$ and the weight 
  $W(z)$ as defined in equations (\ref{eq2}) and (\ref{eq4}).
\label{f19}}
\end{figure}

\begin{figure}
\plotone{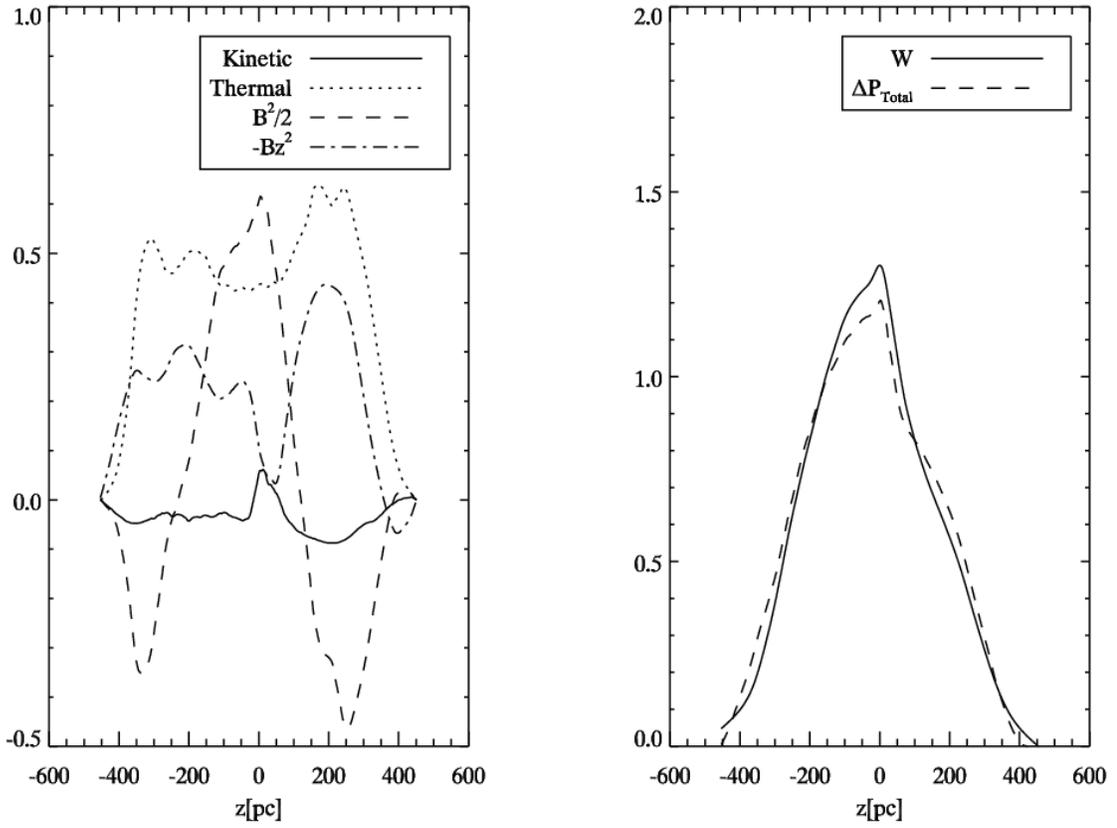}
\caption{The same quantities as in Figure \ref{f19} are shown for the
  mid-gravity model.  The contribution to vertical support from magnetic
  pressure has increased significantly as compared to the other terms.
\label{f20}}
\end{figure}

\begin{figure}
\plotone{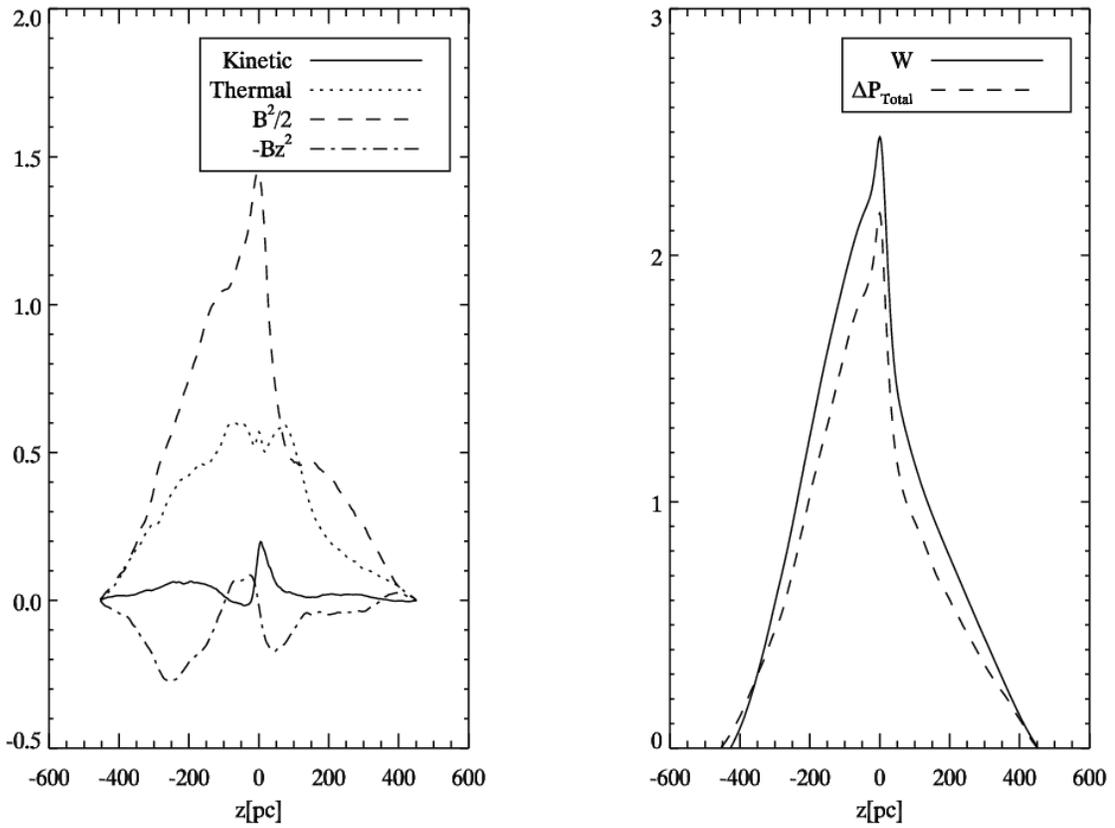}
\caption{The same quantities as in Figure \ref{f19} are shown for the
  high-gravity model.  In the high-gravity model, the magnetic pressure term
  provides most of the vertical support. 
\label{f21}}
\end{figure}

\begin{figure}
\plotone{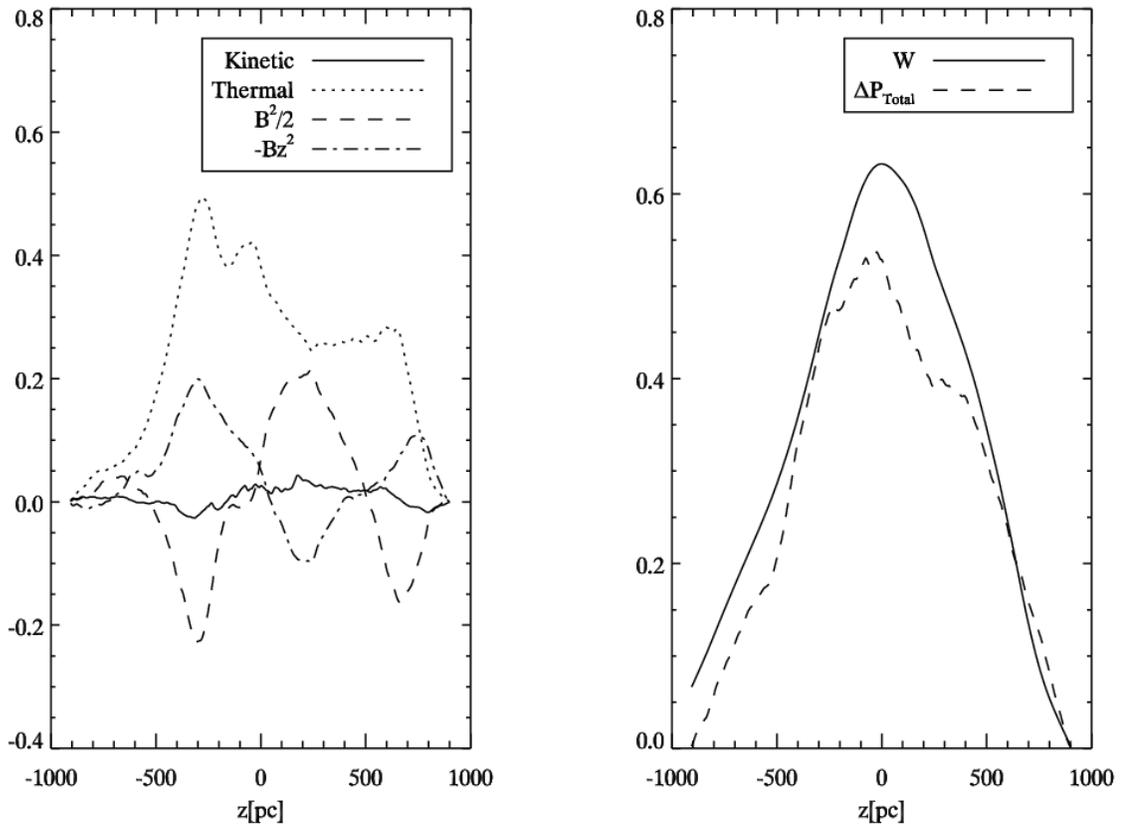}
\caption{The same quantities as in Figure \ref{f19} are shown for the
  outer galaxy model.  In the outer galaxy model the relative
  contribution to vertical support by thermal pressure is large, due
  to the large fraction of warm gas.
\label{f22}}
\end{figure}

In addressing the issue of the vertical distribution of the ISM, and
the relative importance of thermal, kinetic, and magnetic terms to vertical
support, we can test the extent to which quasi-equilibrium is established.
In an equilibrium situation, the left-hand side of equation (\ref{eq1})
would equal zero.  Then, integration of equation (\ref{eq1}) from some 
height $z$ to the top of the box allows us to relate the ``total pressure'' to
the weight of the overlying material.  We define
\begin{equation}
P_{tot}(z) \equiv \langle \rho v_z^2\rangle + \langle P_{th} \rangle +
\frac{\langle(B_x^2 + B_y^2 + B_z^2\rangle}{8\pi} - \frac{\langle
B_z^2\rangle}{4\pi},
\label{eq2}
\end{equation}
with the averages taken over $x$ and $y$ at fixed $z$.  If the time
dependent term in equation (\ref{eq1}) is zero, we would then have
\begin{equation}
\Delta P_{tot}(z)\equiv P_{tot}(z)-P_{tot}(z_{max})= 
\int_z^{z_{max}} (|{ g}_{ext}| \rho)dz'
\equiv W(z).
\label{eq4}
\end{equation}
We can then compare $\Delta P_{tot}(z)$ to the weight $W(z)=
\int_z^{z_{max}} (\tilde g z' \rho)dz'$.  In the right hand panel of
Figure \ref{f19} we plot these two quantities for the low gravity
model, and on the left we plot the four component terms of equation
(\ref{eq2}) for comparison.  The same quantities are shown for the
mid- and high-gravity, and outer galaxy models in Figures \ref{f20},
\ref{f21} and \ref{f22}, respectively.  In each case, these profiles
have been averaged over the last two orbits of the simulation
($t=8-10$ orbits).  Generally the agreement between $W(z)$ and $\Delta
P_{tot}(z)$ is quite good, indicating that the models are indeed in
rough vertical equilibrium.

By comparing the contributions of the four pressure terms in our
different models, we can gain some insight as to what provides support
against the pull of gravity.  An interesting feature of all the models
is that while there are quite large variations in individual $\Delta
P$ components, these compensate each other in such a way that $\Delta
P_{tot}$ is quite smooth.  Another obvious feature, and perhaps one of
the most interesting results, is that the kinetic pressure support
term is small in all of our models.  This suggests that turbulent {\it
  kinetic} pressure driven by the MRI does little to provide vertical
support of the ISM.  Perhaps this is not too surprising because the
total turbulent velocity amplitudes in these models are lower than
those which are observed.  Furthermore, turbulent amplitudes of $v_z$
tend to be smaller than $v_x$ or $v_y$, which further reduces the
effectiveness of the MRI in providing turbulent kinetic support to the
ISM.

In the low gravity model (see Fig. \ref{f19}), all of the $\Delta P$
terms are small compared to the peak values in other models (see Figs.
\ref{f20} - \ref{f22}).  This is because the individual components of
$P_{tot}$ in fact vary relatively weakly with $z$ in the low-gravity
model, so that the $\Delta P$'s are small.  Thus, although
$B^2/(8\pi)$ at any $z$ is larger than each of the other terms by at
least a factor of two in the low gravity model, $\Delta B_z^2/(4\pi)$
and $\Delta P_{th}$ dominate over $\Delta B^2/(8\pi)$.  In the
mid-gravity case, $B^2/(8\pi)$ increases towards small $|z|$ enough so
that, together with the thermal term, $\Delta B^2/(8\pi)$ provides
most of the vertical support near the midplane.  Finally, in the high
gravity case the magnetic pressure term strongly increases inward, and
$\Delta B^2/(8\pi)$ completely dominates the midplane vertical
support.

The thermal term plays an important role in vertical support
in all of the models, and in particular it is the dominant term in the
outer galaxy model.  Here, $\Delta P_{th}$ is larger than all
the other terms (see Fig. \ref{f22}), even though the magnetic pressure
is typically a factor two larger than $P_{th}$; it is the
lack of central concentration in magnetic pressure 
that renders it less important
than the thermal pressure for vertical support.

One last interesting point is that the gas is distributed much more
uniformly in our outer galaxy model (see Fig. \ref{f15}), as compared
to our low gravity model (with the same $\tilde g$) 
with higher surface density.  In comparing
Figures \ref{f19} and \ref{f22}, it appears that the large-scale thermal
pressure gradients are more effective in providing this support in the outer
galaxy model, because of the large fraction of warm gas.

\subsection{Effects of Box Size and Vertical Magnetic Flux}

We have 
performed five additional lower resolution simulations in order to address
the impact of two important parameters, one numerical and the other
physical. 
We first consider the effect of
increasing the size of the simulation domain.  The BOX1, BOX2, and BOX4 models
are identical to our standard model, but performed at half the linear
resolution, and with respective box sizes a factor of 1, 2, and 4 times larger
in the radial and azimuthal directions.  The vertical extent of the domain in
all three BOX models is the same as the standard model, but the resolution is
reduced by a factor of two.  See Table \ref{table1} for a comparison of these
parameters.
The purpose of the BOX tests is to assess whether the saturation
amplitude of the turbulence is sensitive to the size of the domain
simulated.

\begin{figure}
\plotone{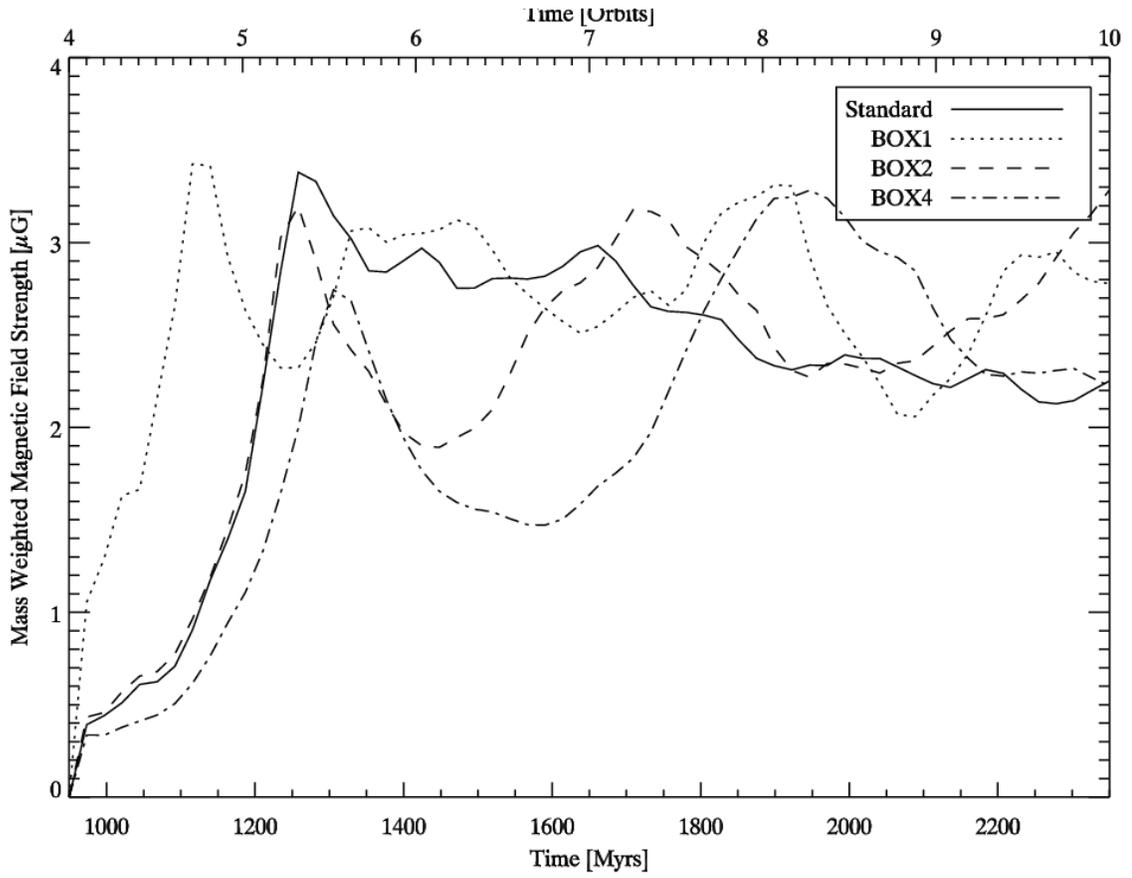}
\caption{Test of saturation amplitude dependence on simulation box size.
\label{f23}}
\end{figure}

The main result of the BOX tests 
is illustrated in Figure \ref{f23}.  We plot the mass
weighted mean magnetic field strength for the warm medium as a function of
time for the standard, BOX1, BOX2, and BOX4 models together.  Averaging over
the final two orbits of these simulations, all three low resolution models
have a similar, but slightly larger, 
time-averaged values of the field strength, $2.6, 2.6,$
and $2.7 \mu{\rm G}$, compared to the mean value $2.3 \mu{\rm G}$ 
for the standard model.  The slight differences in the mean value
may be some indication of a resolution effect on the
saturated state field strength, but longer time averages would be needed to
draw a definitive conclusion. The main conclusion from these tests,
however, is to confirm that the saturated-state value of 
$B^2$ is independent of box size, at least for horizontal domains up
to $8H \times 8 H$.

We have also examined the effect of decreasing the initial vertical
magnetic field strength on the saturated state value of the magnetic field.
The standard value we chose for the mean vertical field, $0.26\mu G$,
is motivated by Galactic observations \citep{1994A&A...288..759H}.
From a general theoretical point of view, however, it is useful to
assess how the value of $|\langle B_z \rangle|$ affects the mean
amplitude of the field that develops.  Previous single-phase models
have found an increase in the saturated-state level according to 
$B^2 \propto |\langle B_z\rangle|$ \citep{1996ApJ...464..690H}
or $B^2 \propto |\langle B_z\rangle|^{3/2}$ \citep{2004ApJ...605..321S}.
Here, two additional simulations, MAG1 and MAG2, were performed with an inital
vertical magnetic field of $0.08 \mu{\rm G}$, corresponding to $\beta=1000$.
The MAG1 model is identical to the BOX1 model, except for the
initial field strength.  The MAG2 model is idential to MAG1, but
we increase the box size by a factor of 2, keeping the linear resolution the
same; thus, MAG2 is a low-$|\langle B_z\rangle|$ version of BOX2.

\begin{figure}
\plotone{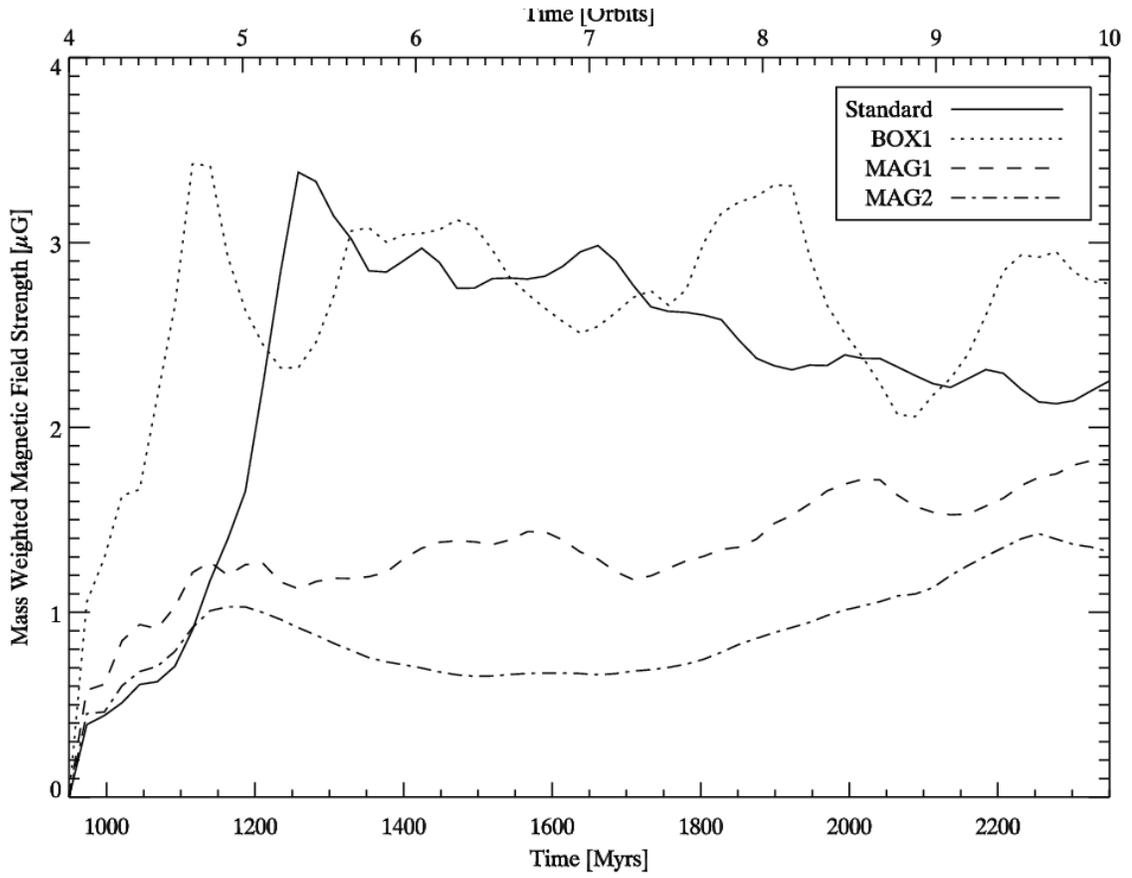}
\caption{Test of saturation amplitude dependence on vertical magnetic flux.
\label{f24}}
\end{figure}

The mass weighted magnetic field strength for the warm medium is shown
for the models MAG1 and MAG2 models in Figure \ref{f24}, with 
the standard and BOX1 models shown for comparison.    
Averaging over the final two orbits, 
$\langle B^2\rangle^{1/2}$ for the MAG1 model is clearly smaller than that
for BOX1, 1.6 and $2.6 \mu{\rm G}$, respectively. In MAG2, 
$\langle B^2\rangle^{1/2}=1.2 \mu{\rm G}$, indicating that 
in this case increasing the box
size also slightly reduces the saturation amplitude.  
The results on lower $\langle B^2\rangle$ 
are generally consistent with the predictions from
previous single-phase models cited above.
We also note, however, that the slow rise in $B^2$ in both MAG1 and MAG2
might continue 
if these models were allowed to run longer in time, yielding a higher final
saturated state value of $B^2$.  Thus, although our tests indicate a
reduction in $\langle B^2\rangle$ if $|\langle B_z\rangle|$ is reduced,
further study would be required for a quantitative analysis.

\section{Summary and Discussion} 
\label{sump3}

\subsection{Model Results}

We have performed numerical simulations of the ISM which include
atomic heating and cooling, galactic shear, magnetic fields and
vertical gravity.  This allows us to study the MRI in the presence of
a two phase medium, with the vertical structure of the disk
determined self-consistently.  Our simulations include three models
with initial surface density typical of that in the solar
neighborhood ($10 \ M_\odot\ {\rm pc}^{-2}$) and vertical gravity
coefficients that vary by a factor 16.
The high gravity value is representative of conditions in the main
star-forming portion of a galactic disk (similar to the Solar
neighborhood), while the lower gravity models allow us to explore how
results vary with the degree of midplane gas concentration.  We have
also performed a fourth simulation with a lower initial surface
density ($6 \ M_\odot\ {\rm pc}^{-2}$) and low gravity, representative
of outer-galaxy conditions.  Our simulations are local, with
$R-\phi-z$ dimensions of $300\times 300\times 900 $ pc for the
inner-disk models, and a box twice as large for the outer-disk model.
Tests with larger numerical boxes suggest that our results are not
sensitive to the size of the  simulation domain; the parameters
adopted enable adequate linear resolution ($\sim 2$ pc) at moderate
computational cost.

Our main results are as follows:

{\it 1.  Thermal phase components -- } For all of our models, and 
similarly to our results in Papers I and II,
we find that thermal instability tends to separate gas into warm and
cold phases on a short timescale, with shear stretching the cold
clouds into filamentary shapes.  Although turbulence maintains a
significant portion of the gas in the thermally-unstable regime, the
density and temperature PDFs still show two distinct -- but broad --
peaks.

For the high-$\Sigma$ models, the cold mass fractions are
(60,60,73)\%, the mass fractions in the unstable temperature range are
(18,16,11)\%, and the warm mass fractions are (22,24,16)\%,
respectively, from low to high gravity.  For comparison, without
turbulence the cold medium makes up (81,86,92)\% from low to high
gravity, and the remaining (19,14,8)\% is in warm gas.  Thus, turbulence
can significantly enhance the amounts of warm and thermally-unstable
gas at the expense of cold gas, and the proportions of gas in the
different phases seem relatively insensitive to the strength of
vertical gravity.  Although the cold mass fraction is somewhat larger
than the $\sim 40\%$ fraction in the local Milky Way estimated
observationally by
\citet{2003ApJ...586.1067H}, our result that warm and unstable gas are
present in similar proportions to each other is consistent with
observational findings.  For our outer-galaxy model, the cold mass
amounts to only $16\%$ of the total, compared to $20\%$ at unstable
temperatures and $ 64\%$ in the warm phase.

{\it 2.  Thermal pressure levels --} The pressure PDFs resulting from 
our simulations are quite broad,
yet the mean midplane pressures in the warm gas are almost identical
($P/k=1500-1600\ \K \cmt$) in all models in spite of the large
differences in the total weight of overlying material (due to varying $g$
and $\Sigma$).  The cold gas is nearly in pressure equilibrium with the warm
gas, with mean midplane thermal pressures of the cold gas within $\sim 20\%$
of the warm-medium values.  Thus, there appears to be a mass exchange
between phases so as to maintain typical thermal pressures comparable to the
geometric mean of $P_{min, cold}$ and $P_{max, warm}$ (here, $800 k\ 
\K \ \cmt$ and $3100 k\  \K \ \cmt$).  This is consistent with
observational indications that the mean local ISM pressure is $P/k
\sim 3000 \ \K \ \cmt$ \citep{2001ApJS..137..297J}, comparable to the
geometric mean between $P_{min, cold}/k=2000 \ \K \ \cmt$ and $P_{max,
  warm}/k=4800 \ \K \ \cmt$ in the standard model of
\citet{1995ApJ...443..152W}.

{\it 3. Vertical Stratification -- } Because the cold clouds have 
densities a factor $\sim 100$ greater
than that in the warm medium, they tend to sink toward the midplane,
creating a differentially-stratified vertical density distribution.
The space-and-time-averaged density profiles show different scale
heights for components of different temperatures.  In particular, the
high-$\Sigma$ inner disk models (with 60-70\% cold gas) all have very
strong central density peaks due to the concentration of cold gas near
the midplane (see Figs. \ref{f14} and \ref{f17}).  
The cold component thicknesses are
only slightly increased relative to the corresponding non-turbulent
models.  For the low-$\Sigma$ outer-disk model, on the other hand, the
overall vertical distribution is somewhat irregular but lacks a
narrow central density peak (see Fig. \ref{f15}).  With a much
smaller fraction of cold gas (20\%), turbulence in the warm medium is
able to maintain a larger cold-disk scale height, $\sim 150\pc$ for
an (approximate) Gaussian fit.  The scale height
$c_s/\sqrt{\tilde{g}}$ for thermally-supported cold gas at mean
temperature $\sim 60 \ \K$ in the outer-disk model would be only $\sim
50\pc$.  Thus, turbulence significantly increases the cold disk's
scale height for this case; the equivalent effective sound speed would
be three times the thermal value.

How do our results compare to observations?  Observations of
high-latitude HI in the Solar neighborhood yield fairly smooth
distributions which can be fit by one or two Gaussians with an overall
effective scale height of $\sim 100-130$ pc
\citep{1990ARA&A..28..215D,1991ApJ...382..182L}.  Cold HI gas is not
included in the fits from high-latitude observations but early
estimates (see \S 1) suggest smaller scale heights than in the warm
gas.  Interestingly, observations of extinction using the 2MASS data
set \citep{2006AA...453..635M} also show a significant central excess
above a profile that follows $\rm sech^2(z/125\pc)$ away from the
midplane, suggestive of a cold component with a scale height much less
than 100 pc.  Recent simulations of the ISM with model supernovae
driving turbulence \citep{2006astro.ph..1005J} have also shown quite 
non-Gaussian vertical density profiles with strong central peaks.

Although the precise scale height of cold gas in the inner Galaxy is
not well known (an updated observational 
study of this would be very valuable), it
is unlikely to be as small as was found for our inner-disk
high-gravity model (a
few 10s of pc).  Thus, we conclude that MRI is not able to account for
the overall vertical HI distribution in the inner portion of the Milky
Way's ISM disk.
When the surface density is high enough that half (or more) of the gas
is forced into the cold component, turbulence driven by MRI in the
diffuse warm gas is not sufficient to lift the cold component off the
midplane.  The gas in the outer Milky Way (and other galaxies) is
observed to have a much larger vertical extent than in the inner disk
(e.g.  \citealt{2000MNRAS.311..361O}), but the detailed vertical
profile is not certain. While our outer-disk model's vertical profile
(see Fig. \ref{f15}) appears reasonable compared to existing observations, 
it is not yet possible to reach a firm conclusion.

{\it 4. Velocity dispersions -- } For all of our models, we find that
the turbulent velocity dispersion decreases with decreasing gas
temperature for the various components.  In the warm phase,
mass-weighted turbulent velocity dispersions are $4-7\ \kms$, while in the cold
phase they are $1-3\ \kms$.  This result differs from that of
Paper II, in which we found similar turbulent velocity dispersions for all
components.  The reason for this difference is well understood.  In
Paper II, we showed that the velocity dispersion in a two-phase medium
depends on the average density, with sufficiently low mean density
allowing velocity dispersion up to $8\ \kms$, and higher mean density
yielding modest velocity dispersions. In the models presented in this
paper, the cold gas tends to sink to the mid-plane, producing
differential stratification.  The mean density is high in the midplane
regions where cold gas is overrepresented, and the velocity dispersion
of cold gas is consequently low.  The mean density is low at high
altitudes where cold gas is depleted, and the velocity dispersion of
warm gas is consequently high (exceeding $10\ \kms$).  This effect is
also self-reinforcing, since the low amplitudes of MRI near the
midplane are not able to drive cold gas to higher latitudes, which
would lower the mean midplane density and raise the turbulent velocity there.

Because the outer-galaxy model has a lower cold mass fraction than the
other models, differential stratification by temperature is less
extreme.  The mean velocity dispersion of cold gas is only a factor
two lower than the warm gas velocity dispersion for this case, while
the ratio is a factor five for the highly-stratified high-gravity
model.  For the inner-galaxy high-$\Sigma$ models, the mean turbulent
velocity dispersions are far lower than the the 7 $\kms$ observed
locally \citep{2003ApJ...586.1067H}.  Thus, other physical processes
must be responsible for driving these large observed velocities.
Even for the outer-galaxy model,
the mean turbulent velocity dispersion is just $\sim 5.5\ \kms$, although
with the high warm fraction the total velocity dispersion including
thermal broadening would be $\approx 8\ \kms$. Turbulent and thermal
velocity dispersions cannot be separated in HI emission observations
of external galaxies, so it is not yet known what the true turbulent
amplitudes (or warm/cold mass fractions) in the outer portions of
galaxies are (see \S 4.2).

{\it 5. Magnetic field strengths --} An interesting result of this
work, in agreement with the results of the unstratified models of
Paper II, is that the saturated-state magnetic energy density is
fairly independent of mean density (i.e. the proportions of warm and
cold gas).  While there is a factor 30
difference in the mean midplane density between the high-gravity,
high-$\Sigma$ inner disk model and the low-gravity, low-$\Sigma$ outer
disk model ($\bar n=13 \ \cmt$ vs.  $\bar n=0.44 \ \cmt$), the ratio of
magnetic energy densities is only a factor two ($\beta=0.3$ vs.
$\beta=0.6$).  This suggests that MRI saturates when the magnetic
energy density is in approximate equipartition with thermal energy.
While it is important to understand this process in detail, one
interpretation of this result might be that growth of the magnetic
field (at a rate $\sim \Omega \beta^{1/2}$ when $B_y$ dominates) is
balanced by magnetic reconnection (at a rate $\sim \Omega$ or $\sim
\Omega \beta^{-1/2}$ depending on vertical support) when $\beta \sim
1$ (see also \citet{2001ApJ...561L.179S, 2004ApJ...605..321S}, who
consider saturation levels in MRI models with explicit resistivity).

The range in midplane values of the plasma parameter $\beta=0.3-0.6$ for all
of the models performed, including both warm and cold phases of gas.  For our
chosen cooling curve, which yields midplane thermal pressure $P/k=1500-1600 \ 
\K \ \cmt$, this corresponds to midplane magnetic field strengths of $3-4 \ 
\mu{\rm G}$.  Averaging over all of the gas, the saturated-state magnetic
field strengths range from $2-4 \ \mu{\rm G}$, lowest for the warm gas in the
outer-galaxy model, and highest for the cold gas in the high-gravity model.
Allowing for the fact that our adopted cooling curve results in mean thermal
pressure lower by a factor $\sim 2$ compared to Solar-neighborhood
observations, these magnetic field strengths are entirely consistent with
observed local Milky Way magentic fields of $\approx 6 \ \mu{\rm G}$
\citep{2005ApJ...624..773H}, and similar values in external galaxies
\citep{2006astro.ph..3531B}.  This shows that the MRI is capable of amplifying
weak magnetic fields up to realistic interstellar values, and as such is one
of the most important elements of the galactic dynamo.  We note that 
the quantitative levels above were obtained using $\langle
B_z\rangle=0.26 \mu$G, motivated by observations.  Models
with smaller $\langle B_z\rangle$ result in lower 
saturated-state $B^2$, consistent with the results of single-phase MRI
simulations.

{\it 6. Vertical support of the disk --}  Analysis of the
contributions to the time-averaged momentum equation show
that our models reach approximate vertical equilibrium between the
downward force of gravity and the upward forces of the combined
effects of thermal pressure, kinetic pressure, and magnetic pressure
and tension.  The dominant terms are thermal and magnetic pressure
gradients; the latter is most important in regions where the majority of the
mass is in cold gas, while the latter predominates elsewhere.

While the profiles of weight and total pressure difference $\Delta
P=P(z)-P(z_{max})$ are both fairly smooth as a function of $z$, the
individual terms in $\Delta P$ can be quite irregular.  In addition,
even when there is strongly dominant support by a single term, as for
the magnetic-pressure supported midplane regions of the high-gravity case,
the vertical variation of magnetic pressure and gas density 
differ, such that $H=(P_B/(\rho \tilde g))^{1/2}$ at the midplane 
does not yield an accurate value for the scale height. In this case,
it overestimates the density scale height of 12 pc by about 60\%.

\subsection{Astronomical Persectives}

A major motivation for the present study, together with Papers I and
II, has been to quantify the amplitude of turbulent motions that can
be driven by MRI in interstellar gas with realistic properties.
\citet{1999ApJ...511..660S} originally proposed that the observed
near-constant HI velocity dispersion $\approx 6\ \kms$ reported in the
outer parts of NGC 1058 \citep{1990ApJ...352..522D} could potentially
be attributed to random motions of an ensemble of small cold clouds
associated with MRI-driven turbulence.  In fact, we find that the
saturated-state velocity dispersion in the cold-cloud ensemble is
always fairly low, so that the cold medium's intrinsic contribution to
the total (thermal+nonthermal averaged over all phases) 
velocity dispersion would exceed the
contribution from the warm medium only if the warm medium represents a
miniscule ($<5\%$) proportion of the gas mass.  Since MRI-driven
turbulence even in the warm medium is at most sonic and the vertical
component is smaller than the in-plane components, thermal broadening
should in practice exceed turbulent broadening for face-on galaxies.

Thus, we suggest that observed values of the velocity dispersion lower
than $\approx 8\ \kms$ may imply the presence of cold gas with
velocity width smaller than the 21 cm receiver's channels.  This could
have the effect of reducing the inferred linewidth relative to what
would be measured if only warm gas were present.  For example, if warm
medium with thermal dispersion of $8\ \kms$ makes up 40\% of the total
mass  and the intrinsic instrumental linewidth and turbulent velocity
dispersion are each $3\ \kms$, then adding in quadrature would yield a total
linewidth of 6.6 $\kms$.  This is quite similar to the mean value
reported for NGC 1068.  If, on the other hand, the observed velocity
dispersion exceeds $\sim 8 \ \kms$, we suggest that there must be 
little cold gas.  MRI-driven turbulence in the warm medium can
contribute at a moderate level to the velocity dispersion in this
case.  However, it is possible that corrugation in the disk is
allowing rotation to contaminate the inferred velocity dispersion when values
as large as 15 $\kms$ are reported in regions where there is little
observed star formation (and negligible spiral structure) in external galaxies.

Another important astronomical motivation for our work has been to
help understand what controls the spatial extent of disk star
formation.  Usually, the sharp drop in star formation in the outer
parts of disks is attributed to a gravitational threshold; if the
effective velocity dispersion is $c_{\rm eff} = 6 \ \kms$, the
mean critical Toomre parameter $\kappa c_{\rm eff}/(\pi G \Sigma)$ is
measured to be 1.4 at the threshold for active star formation
\citep{1989ApJ...344..685K,2001ApJ...555..301M,2003MNRAS.346.1215B}.
The value $Q_{\rm crit}=1.4$ is consistent with theoretical measures
of the $Q$-threshold
for nonlinear gravitational runaway, allowing for both
self-gravitating gas and stars \citep{2007astro.ph..1755K}.

An alternative proposal is that the star formation threshold simply
marks the radius in the disk inside which the midplane hydrostatic
pressure is sufficient for a cold phase to be present
\citep{2004ApJ...609..667S}. This radius depends through the heating
and cooling curves on the metallicity and level of ambient UV radiation
present, but since $\Sigma=[2 P/(\pi G)]^{1/2}$ for a
non-turbulent, pure-gas disk and $P_{\rm min,cold}/k=200-600\ \K \ \cmt$
in the outer parts of disks
\citep{1994ApJ...435L.121E,2003ApJ...587..278W}, this would typically
be near where $\Sigma=2-4 \ \Msun \pc^{-2}$.  \footnote{Since MRI can
compress gas to pressures higher than would be possible from
hydrostatic equilibrium alone, cold gas would in reality be present even
at lower values of the total surface density.}
For a flat rotation curve with circular velocity $v_{\rm c}$, the
value of $Q$ at radius $R$ is
\begin{equation}
Q=1.4 \left(\frac{c_{\rm eff}}{6\, \kms }\right)
      \left(\frac{v_{\rm c}}{200\, \kms}\right)
      \left(\frac{R}{15\, \kpc}\right)^{-1}
\left(\frac{\Sigma}{6 \Msun \pc^{-2}}\right)^{-1}.
\label{Qeq}
\end{equation}
Schaye's proposal is based on the idea that if $c_{\rm eff}$ were {\it
  only} to include the thermal sound speed $c_s$, then with $c_s \sim
1\ \kms$ for cold gas, gravitational instability and hence active star
formation would develop essentially wherever there is a cold component 
(provided that the not-very-restrictive condition
$R \Sigma_{cold}>
15 \ {\rm kpc} \Msun {\rm pc }^{-2} \times
[v_{\rm c}/200\, \kms] $ is met).

The proposal of Schaye in fact does not appear consistent with
observations, since observed star formation thresholds often occur
where the ISM is predominantly molecular, well inside the maximum
radius for cold HI to be present\citep{2001ApJ...555..301M}.
\footnote{\citet{2006AJ....131..363D} also give direct evidence that
  cold gas is present in the outer, non-star-forming parts of NGC
  6822, based on the clear broad/narrow components of
  high-velocity-resolution HI line profiles.}  But if cold gas is
present in the outer parts of galaxies, then its thermal pressure is
certainly inadequate to prevent gravitational instability.  This
implies that $c_{\rm eff}$ must include other terms that are large
enough to yield $Q_{\rm cold}>Q_{\rm crit}\sim 1.4$.  We have shown
that turbulent velocities of $\sim 2.5\ \kms$ develop in the cold
medium under outer-disk conditions where the cold mass fraction is
small; even if $\Sigma_{cold}$ is as large as $2 \ \Msun {\rm pc}^{-2}$
this turbulence would suppress instability in the cold medium out
beyond $20$ kpc.  Perhaps even more important, however, are the
magnetic fields.  

Our models show that MRI amplifies the magnetic field until the
midplane magnetic pressure is $\sim 2-3$ times the thermal pressure
$P/k\approx 1500 \ \K \ \cmt$.
If the mean midplane density is $0.5\ \cmt$ or less (as is true for our
outer-galaxy model), then the Alfv\'en speed $v_{\rm A}= B/\sqrt{4 \pi
  \bar\rho}$ at the midplane would exceed $8\ \kms$.  Even if the
midplane thermal pressure is somewhat lower in the outer parts of
galaxies than for our models, the magnetic fields would still be quite
large. {\it If} the substitution $c_{\rm eff}\rightarrow v_{\rm A}$
in equation (\ref{Qeq}) legitimately describes the effect of turbulent
magnetic fields on gravitational instability in a two-phase ISM disk,
then it may be primarily MRI-driven magnetic fields that suppress
outer-disk star formation. \footnote{MRI could not on its own suppress
  star formation in the inner disk, however, since cold gas settles at
  the midplane when too much is present, resulting in a high $\bar
  \rho$ and low $v_{\rm A}$ at $z=0$.}  Whether this naive approach is
even approximately valid or not, however, is not certain.  In some
circumstances magnetic fields are known to suppress self-gravitational
instability, and we have seen here that they provide support against
vertical gravity in the simulations we have performed.  In other
situations, however, turbulent magnetic fields can encourage disk
instabilities by transferring angular momentum out of condensations
(e.g. \citealt{2003ApJ...599.1157K}).  Addressing this question
directly with numerical simulations represents an important
problem for future study.

\acknowledgments 

We are grateful to Jay Lockman, Chris McKee, and Tom Troland for
valuable discussions. This work was supported in part by grants AST
0205972 and AST 0507315 from the National Science Foundation.  Most of
the simulations presented here were performed on the Thunderhead
cluster at Goddard Space Flight Center, and some were performed on the
CTC cluster in the UMD Department of Astronomy. This research has made
use of NASA's Astrophysics Data System.

\bibliographystyle{apj}
\bibliography{master}

\begin{thebibliography}{53}
\expandafter\ifx\csname natexlab\endcsname\relax\def\natexlab#1{#1}\fi

\bibitem[{{Baker} \& {Burton}(1975)}]{1975ApJ...198..281B}
{Baker}, P.~L., \& {Burton}, W.~B. 1975, \apj, 198, 281

\bibitem[{{Balbus}(2003)}]{2003ARA&A..41..555B}
{Balbus}, S.~A. 2003, \araa, 41, 555

\bibitem[{{Balbus} \& {Hawley}(1991)}]{1991ApJ...376..214B}
{Balbus}, S.~A., \& {Hawley}, J.~F. 1991, \apj, 376, 214

\bibitem[{{Beck}(2006)}]{2006astro.ph..3531B}
{Beck}, R. 2006, ArXiv Astrophysics e-prints

\bibitem[{{Boissier} {et~al.}(2003){Boissier}, {Prantzos}, {Boselli}, \&
  {Gavazzi}}]{2003MNRAS.346.1215B}
{Boissier}, S., {Prantzos}, N., {Boselli}, A., \& {Gavazzi}, G. 2003, \mnras,
  346, 1215

\bibitem[{{Boulares} \& {Cox}(1990)}]{1990ApJ...365..544B}
{Boulares}, A., \& {Cox}, D.~P. 1990, \apj, 365, 544

\bibitem[{{de Avillez} \& {Breitschwerdt}(2004)}]{2004A&A...425..899D}
{de Avillez}, M.~A., \& {Breitschwerdt}, D. 2004, \aap, 425, 899

\bibitem[{{de Avillez} \& {Breitschwerdt}(2005)}]{2005A&A...436..585D}
---. 2005, \aap, 436, 585

\bibitem[{{de Blok} \& {Walter}(2006)}]{2006AJ....131..363D}
{de Blok}, W.~J.~G., \& {Walter}, F. 2006, \aj, 131, 363

\bibitem[{{Dib} \& {Burkert}(2005)}]{2005ApJ...630..238D}
{Dib}, S., \& {Burkert}, A. 2005, \apj, 630, 238

\bibitem[{{Dickey} {et~al.}(1990){Dickey}, {Hanson}, \&
  {Helou}}]{1990ApJ...352..522D}
{Dickey}, J.~M., {Hanson}, M.~M., \& {Helou}, G. 1990, \apj, 352, 522

\bibitem[{{Dickey} \& {Lockman}(1990)}]{1990ARA&A..28..215D}
{Dickey}, J.~M., \& {Lockman}, F.~J. 1990, \araa, 28, 215

\bibitem[{{Dziourkevitch} {et~al.}(2004){Dziourkevitch}, {Elstner}, \&
  {R{\"u}diger}}]{2004A&A...423L..29D}
{Dziourkevitch}, N., {Elstner}, D., \& {R{\"u}diger}, G. 2004, \aap, 423, L29

\bibitem[{{Elmegreen} \& {Parravano}(1994)}]{1994ApJ...435L.121E}
{Elmegreen}, B.~G., \& {Parravano}, A. 1994, \apjl, 435, L121+

\bibitem[{{Falgarone} \& {Lequeux}(1973)}]{1973A&A....25..253F}
{Falgarone}, E., \& {Lequeux}, J. 1973, \aap, 25, 253

\bibitem[{{Field} {et~al.}(1969){Field}, {Goldsmith}, \&
  {Habing}}]{1969ApJ...155L.149F}
{Field}, G.~B., {Goldsmith}, D.~W., \& {Habing}, H.~J. 1969, \apjl, 155, L149+

\bibitem[{{Fricke}(1969)}]{1969A&A.....1..388F}
{Fricke}, K. 1969, \aap, 1, 388

\bibitem[{{Han} \& {Qiao}(1994)}]{1994A&A...288..759H}
{Han}, J.~L., \& {Qiao}, G.~J. 1994, \aap, 288, 759

\bibitem[{{Hawley} \& {Balbus}(1991)}]{1991ApJ...376..223H}
{Hawley}, J.~F., \& {Balbus}, S.~A. 1991, \apj, 376, 223

\bibitem[{{Hawley} \& {Balbus}(1992)}]{1992ApJ...400..595H}
---. 1992, \apj, 400, 595

\bibitem[{{Hawley} {et~al.}(1995){Hawley}, {Gammie}, \&
  {Balbus}}]{1995ApJ...440..742H}
{Hawley}, J.~F., {Gammie}, C.~F., \& {Balbus}, S.~A. 1995, \apj, 440, 742

\bibitem[{{Hawley} {et~al.}(1996){Hawley}, {Gammie}, \&
  {Balbus}}]{1996ApJ...464..690H}
---. 1996, \apj, 464, 690

\bibitem[{{Heiles} \& {Troland}(2003)}]{2003ApJ...586.1067H}
{Heiles}, C., \& {Troland}, T.~H. 2003, \apj, 586, 1067

\bibitem[{{Heiles} \& {Troland}(2005)}]{2005ApJ...624..773H}
---. 2005, \apj, 624, 773

\bibitem[{{Holmberg} \& {Flynn}(2000)}]{2000MNRAS.313..209H}
{Holmberg}, J., \& {Flynn}, C. 2000, \mnras, 313, 209

\bibitem[{{Jenkins} \& {Tripp}(2001)}]{2001ApJS..137..297J}
{Jenkins}, E.~B., \& {Tripp}, T.~M. 2001, \apjs, 137, 297

\bibitem[{{Joung} \& {Mac Low}(2006)}]{2006astro.ph..1005J}
{Joung}, M.~K.~R., \& {Mac Low}, M.-M. 2006, ArXiv Astrophysics e-prints

\bibitem[{{Kennicutt}(1989)}]{1989ApJ...344..685K}
{Kennicutt}, Jr., R.~C. 1989, \apj, 344, 685

\bibitem[{{Kim} \& {Ostriker}(2007)}]{2007astro.ph..1755K}
{Kim}, W.-T., \& {Ostriker}, E.~C. 2007, ArXiv Astrophysics e-prints

\bibitem[{{Kim} {et~al.}(2003){Kim}, {Ostriker}, \&
  {Stone}}]{2003ApJ...599.1157K}
{Kim}, W.-T., {Ostriker}, E.~C., \& {Stone}, J.~M. 2003, \apj, 599, 1157

\bibitem[{{Lockman} \& {Gehman}(1991)}]{1991ApJ...382..182L}
{Lockman}, F.~J., \& {Gehman}, C.~S. 1991, \apj, 382, 182

\bibitem[{{Marshall} {et~al.}(2006){Marshall}, {Robin}, {Reyl{\'e}},
  {Schultheis}, \& {Picaud}}]{2006AA...453..635M}
{Marshall}, D.~J., {Robin}, A.~C., {Reyl{\'e}}, C., {Schultheis}, M., \&
  {Picaud}, S. 2006, \aap, 453, 635

\bibitem[{{Martin} \& {Kennicutt}(2001)}]{2001ApJ...555..301M}
{Martin}, C.~L., \& {Kennicutt}, Jr., R.~C. 2001, \apj, 555, 301

\bibitem[{{McKee} \& {Ostriker}(1977)}]{1977ApJ...218..148M}
{McKee}, C.~F., \& {Ostriker}, J.~P. 1977, \apj, 218, 148

\bibitem[{{Miller} \& {Stone}(2000)}]{2000ApJ...534..398M}
{Miller}, K.~A., \& {Stone}, J.~M. 2000, \apj, 534, 398

\bibitem[{{Mohan} {et~al.}(2004){Mohan}, {Dwarakanath}, \&
  {Srinivasan}}]{2004JApA...25..185M}
{Mohan}, R., {Dwarakanath}, K.~S., \& {Srinivasan}, G. 2004, Journal of
  Astrophysics and Astronomy, 25, 185

\bibitem[{{Olling} \& {Merrifield}(2000)}]{2000MNRAS.311..361O}
{Olling}, R.~P., \& {Merrifield}, M.~R. 2000, \mnras, 311, 361

\bibitem[{{Piontek} \& {Ostriker}(2004)}]{2004ApJ...601..905P}
{Piontek}, R.~A., \& {Ostriker}, E.~C. 2004, \apj, 601, 905

\bibitem[{{Piontek} \& {Ostriker}(2005)}]{2005ApJ...629..849P}
---. 2005, \apj, 629, 849

\bibitem[{{S{\'a}nchez-Salcedo} {et~al.}(2002){S{\'a}nchez-Salcedo},
  {V{\'a}zquez-Semadeni}, \& {Gazol}}]{2002ApJ...577..768S}
{S{\'a}nchez-Salcedo}, F.~J., {V{\'a}zquez-Semadeni}, E., \& {Gazol}, A. 2002,
  \apj, 577, 768

\bibitem[{{Sano} \& {Inutsuka}(2001)}]{2001ApJ...561L.179S}
{Sano}, T., \& {Inutsuka}, S.-i. 2001, \apjl, 561, L179

\bibitem[{{Sano} {et~al.}(2004){Sano}, {Inutsuka}, {Turner}, \&
  {Stone}}]{2004ApJ...605..321S}
{Sano}, T., {Inutsuka}, S.-i., {Turner}, N.~J., \& {Stone}, J.~M. 2004, \apj,
  605, 321

\bibitem[{{Schaye}(2004)}]{2004ApJ...609..667S}
{Schaye}, J. 2004, \apj, 609, 667

\bibitem[{{Sellwood} \& {Balbus}(1999)}]{1999ApJ...511..660S}
{Sellwood}, J.~A., \& {Balbus}, S.~A. 1999, \apj, 511, 660

\bibitem[{{Slyz} {et~al.}(2005){Slyz}, {Devriendt}, {Bryan}, \&
  {Silk}}]{2005MNRAS.356..737S}
{Slyz}, A.~D., {Devriendt}, J.~E.~G., {Bryan}, G., \& {Silk}, J. 2005, \mnras,
  356, 737

\bibitem[{{Spitzer}(1978)}]{1978ppim.book.....S}
{Spitzer}, L. 1978, {Physical processes in the interstellar medium} (New York
  Wiley-Interscience, 1978.~333 p.)

\bibitem[{{Stone} {et~al.}(2000){Stone}, {Gammie}, {Balbus}, \&
  {Hawley}}]{2000prpl.conf..589S}
{Stone}, J.~M., {Gammie}, C.~F., {Balbus}, S.~A., \& {Hawley}, J.~F. 2000,
  Protostars and Planets IV, 589

\bibitem[{{Stone} {et~al.}(1996){Stone}, {Hawley}, {Gammie}, \&
  {Balbus}}]{1996ApJ...463..656S}
{Stone}, J.~M., {Hawley}, J.~F., {Gammie}, C.~F., \& {Balbus}, S.~A. 1996,
  \apj, 463, 656

\bibitem[{{Stone} \& {Norman}(1992{\natexlab{a}})}]{1992ApJS...80..753S}
{Stone}, J.~M., \& {Norman}, M.~L. 1992{\natexlab{a}}, \apjs, 80, 753

\bibitem[{{Stone} \& {Norman}(1992{\natexlab{b}})}]{1992ApJS...80..791S}
---. 1992{\natexlab{b}}, \apjs, 80, 791

\bibitem[{{van Zee} \& {Bryant}(1999)}]{1999AJ....118.2172V}
{van Zee}, L., \& {Bryant}, J. 1999, \aj, 118, 2172

\bibitem[{{Wolfire} {et~al.}(1995){Wolfire}, {Hollenbach}, {McKee}, {Tielens},
  \& {Bakes}}]{1995ApJ...443..152W}
{Wolfire}, M.~G., {Hollenbach}, D., {McKee}, C.~F., {Tielens}, A.~G.~G.~M., \&
  {Bakes}, E.~L.~O. 1995, \apj, 443, 152

\bibitem[{{Wolfire} {et~al.}(2003){Wolfire}, {McKee}, {Hollenbach}, \&
  {Tielens}}]{2003ApJ...587..278W}
{Wolfire}, M.~G., {McKee}, C.~F., {Hollenbach}, D., \& {Tielens}, A.~G.~G.~M.
  2003, \apj, 587, 278

\end{thebibliography}


\clearpage

\begin{deluxetable}{lccccccccc}
\tabletypesize{\scriptsize}
\tablecaption{Model parameters for simulations performed \label{table1}}
\tablewidth{0pt}
\tablehead{
\colhead{Model} & \colhead{$\tilde{g} \ [10^{-31} {\rm s}^{-2}]$} &
\colhead{$H \ [pc]$} & \colhead{Boxsize [$H$]} & $N_x \times N_y \times N_z$ & \colhead{$T_o \ [K]$} &
\colhead{$\rho_o \ [cm^{-3}]$} & \colhead{$\Sigma \ [\Msun \pc^{-2}]$} &
\colhead{$B_z \ [\mu G]$}
}
\startdata
Low Gravity  &1.94 &150 &$2\times 2\times 6$ &$128\times 128\times 384$ &600   &0.85  &10 & 0.26 \\
Standard     &7.76 &150 &$2\times 2\times 6$ &$128\times 128\times 384$ &2500  &0.85  &10 & 0.26 \\
High Gravity &31.0 &150 &$2\times 2\times 6$ &$128\times 128\times 384$ &10000 &0.85  &10 & 0.26 \\
Outer Galaxy &1.94 &300 &$2\times 2\times 6$ &$128\times 128\times 384$ &2500  &0.26  &6  & 0.26 \\
BOX1         &7.76 &150 &$2\times 2\times 6$ &$64\times 64\times 192$ &2500  &0.85  &10 & 0.26 \\
BOX2         &7.76 &150 &$4\times 4\times 6$ &$128\times 128\times 192$ &2500  &0.85  &10 & 0.26 \\
BOX3         &7.76 &150 &$8\times 8\times 6$ &$256\times 256\times 192$ &2500  &0.85  &10 & 0.26 \\
MAG1         &7.76 &150 &$2\times 2\times 6$ &$64\times 64\times 192$ &2500  &0.85  &10 & 0.08 \\
MAG2         &7.76 &150 &$4\times 4\times 6$ &$128\times 128\times 192$ &2500  &0.85  &10 & 0.08 \\
\enddata

\end{deluxetable}

\clearpage 

\clearpage 

\clearpage 

\clearpage 

\clearpage 

\clearpage 

\clearpage 

\clearpage 

\clearpage 

\clearpage 

\clearpage 

\clearpage 

\clearpage 

\clearpage 

\clearpage 

\clearpage 

\clearpage 

\clearpage 

\clearpage 

\clearpage 

\clearpage 

\clearpage

\clearpage

\clearpage

\end{document}